\documentclass[twocolumn,twocolappendix]{aastex63}

\usepackage{textcomp}
\usepackage{array,multirow}
\usepackage{natbib}
\usepackage{amsmath}
\begin{document}

\title{Measuring Supermassive Black Hole Spin in the \textit{Chandra} COSMOS-Legacy Survey}

\author{Mackenzie L. Jones}
\affil{Center for Astrophysics | Harvard \& Smithsonian, 60 Garden St, Cambridge, MA 02138, USA}

\author{Laura Brenneman}
\affil{Center for Astrophysics | Harvard \& Smithsonian, 60 Garden St, Cambridge, MA 02138, USA}

\author{Francesca Civano}
\affil{Center for Astrophysics | Harvard \& Smithsonian, 60 Garden St, Cambridge, MA 02138, USA}

\author{Giorgio Lanzuisi}
\affil{INAF - Osservatorio di Astrofisica e Scienza dello Spazio di Bologna, Via Piero Gobetti, 93/3, 40129, Bologna, Italy}

\author{Stefano Marchesi}
\affil{INAF - Osservatorio di Astrofisica e Scienza dello Spazio di Bologna, Via Piero Gobetti, 93/3, 40129, Bologna, Italy}
\affil{Department of Physics and Astronomy, Clemson University,  Kinard Lab of Physics, Clemson, SC 29634, USA}

\begin{abstract}

Spin measurements of supermassive black holes (SMBHs) provide crucial constraints on the accretion processes that power active galactic nuclei (AGN), fuel outflows, and trigger black hole growth. 
However, spin measurements are mainly limited to a few dozen nearby sources for which high quality, high S/N spectra (e.g., from \textit{Chandra}, \textit{XMM-Newton}, \textit{Suzaku}, \textit{NuSTAR}) are available. Here we measure the average SMBH spin of $\sim$1900 AGN in the \textit{Chandra} COSMOS-Legacy survey using spectral stacking analysis. We find broad Fe K$\alpha$ line emission in the average COSMOS spectrum (Gaussian width $\sigma=0.27\pm0.05$ keV), and by fitting this emission line profile with relativistic line models, we measure the average black hole spin parameter $a=0.62~\substack{+0.07 \\ -0.17}$. The sample size, availability of multiwavelength data, and spatial resolution of the COSMOS Legacy field also provide a unique environment to investigate the average SMBH spin as a function of other observables (e.g., redshift, luminosity) up to $z\sim5.3$. We find that optically classified Type 1 sources have broader Fe K$\alpha$ line emission than Type 2 sources. X-ray unobscured and obscured sources, as defined by their column densities, have widths that are consistent with the optically defined unobscured and obscured sources, respectively. There is some evidence for evolution of the Fe K$\alpha$ width and black hole spin parameter with luminosity, but not conclusively with redshift.
The results of this work provide insights into the average spins of SMBHs in AGN, shedding light on their growth mechanisms and observed co-evolution with their host galaxies.

\end{abstract}

\keywords{galaxies: active; X-rays: galaxies}

\section{Introduction}\label{sec:intro}

Supermassive black holes (SMBHs) are found nearly ubiquitously in the centers of galaxies across cosmic time. As they grow via mass accretion in one of the most efficient engines in the universe, they emit radiation as active galactic nuclei (AGN). 
These powerful outflows shape the surrounding interstellar and intergalactic media (e.g., \citealt{Fab03,Fab06,Hop10,Hop12}), although the exact physical mechanisms of this process are not yet fully understood (e.g., see \citealt{Mor17} for a review). 

It is known, however, that the spins of SMBHs, and by extension the energy imparted on the surrounding environment, must be a critical parameter of these outflows. 
Furthermore, SMBH spins are impacted by mergers and accretion events, leaving a ``fossil record'' of black hole formation (\citealt{Rey19}) and providing compelling evidence of black hole-galaxy coevolution (e.g., \citealt{Ber08}).

Despite its importance, measuring SMBH spin magnitudes and directions has only been possible in the past decade with the development of new theoretical models coupled with high quality, high signal-to-noise (S/N) spectra (e.g., from \textit{Chandra}, \textit{XMM-Newton}, \textit{Suzaku}, \textit{NuSTAR}) of nearby AGN (e.g., \citealt{Dov04,Bre06,Dau10,Dau13,Gar14}).

Signatures of a relativistically spinning black hole are imparted onto the characteristic features of the X-ray ``reflection spectrum'' (e.g., \citealt{Fab89,Lao91,Geo91}) via Special and General relativistic effects. 
The most prominent feature of this reflection spectrum is the neutral Fe K$\alpha$ emission line at a rest-frame energy of $E = 6.4$ keV (e.g., \citealt{Bre06}). 
In order to constrain BH spin with relativistic line models, spectra typically must have $>200,000$ counts over the $2-10$ keV bandpass in order to obtain adequate S/N to distinguish the reflection features from those of the continuum or line-of-sight absorption intrinsic to the AGN system (e.g., \citealt{Gua06}).

Since the first SMBH spin measurements were reported by \citet{Bre06} for MCG--6-30-15, there have only been a few dozen spins measured for individual SMBHs (e.g., \citealt{Bre09,Rey14,Vas16}; see also \citealt{Bre13,Rey19} for reviews). These spins are almost certainly biased toward high, prograde spin values because of selection effects: the sample of AGN from which they are drawn represent the brightest and/or closest sources (e.g., \citealt{Bre11,Rey12,Bon16,Vas16,Fab19}).

We are currently limited by S/N requirements when probing the black hole spin population using X-ray reflection-based spin measurements for AGN with lower fluxes and at greater distances. 
Gravitationally lensed systems, however, provide a unique environment to measure spins at higher redshifts by capitalizing on the magnification imparted from the lens which boosts the observed flux of the background AGN. Measuring SMBH spin via gravitational lensing requires \textit{Chandra}'s spectral resolving power and unparalleled spatial resolution.
In particular, \citet{Rei14} successfully measured the spin of a lensed quasar at $z=0.658$ to be $a>0.66$ at the 5$\sigma$ level. Similarly, \citet{Rey14} found $a=0.74\pm0.06$ at 90\% confidence for the central SMBH in the Einstein Cross, and \citet{Dai19} jointly fit four quasars to find $a=0.8\pm0.16$.

Individual measurements using gravitational lensing can still be restricted by S/N requirements, so stacking methods have been used to overcome this limitation to varying success, using dozens to thousands of low-count AGN spectra from AGN surveys: \textit{XMM} (e.g., \citealt{Cor08,Iwa12,Fal13,Fal14,Liu16}), \textit{NuSTAR} (e.g., \citealt{Del17,Zap18}), and \textit{Chandra} (e.g., \citealt{Fal12}). 
Despite the higher collecting area of \textit{XMM} compared to that of \textit{NuSTAR} or \textit{Chandra}, spectral stacking of \textit{XMM} sources have not yet yielded conclusive measurements of the average SMBH spins. 
\citet{Cor08} stacked $\sim$600 Type-1 AGN spectra from the \textit{XMM} medium survey, \citet{Iwa12} examined $\sim$1000 AGN spectra from \textit{XMM}-COSMOS, \citet{Cha12} selected 248 AGN spectra between $1<z<5$ from the \textit{2XMM} catalog, and \citet{Liu16} stacked 2512 AGN spectra in the \textit{XMM}-XXL North survey; none of these analyses found evidence of broad Fe K$\alpha$ emission. 
Similarly, \citet{Fal13} did not detect relativistic broadening of the Fe K$\alpha$ emission line in the \textit{XMM} CDF-S. However, by stacking 263 \textit{XMM}-selected unabsorbed AGN spectra from the V\`eron-Cetty and V\`eron catalogs, \citet{Fal14} found that the addition of a relativistic line profile to a narrow Fe K$\alpha$ emission line profile improved the spectral fit of their stacked AGN spectra at the 6$\sigma$ level, but were unable to constrain an average black hole spin.
Likewise, \citealt{Wal15} stacked the spectra of 27 lensed quasars to investigate spins in SMBHs at $z\sim1-4$, and were able to constrain the average black hole spin in their sample, measuring an average spin of $a\sim0.7$.

These works represent important first steps toward measuring black hole spin at high redshifts. In this work, we have built upon this progress by leveraging the population of $\sim1800$ AGN in the \textit{Chandra}-COSMOS Legacy survey (hereafter, CCLS; \citealt{Civ16,Mar16O,Mar16X}).
We investigate the average black hole spin of the AGN population probed by the CCLS, including possible dependencies on redshift and environmental factors, by performing a stacking analysis of thousands of low-count spectra (similar to that of \citealt{Wal15}), but also as a function of redshift out to $z\sim5.3$. 

We describe the CCLS and our source selection in Section \ref{sec:obs}. Our stacking analysis and the fitting of the Fe K$\alpha$ emission are described in Sections \ref{sec:stack} and \ref{sec:fit}, respectively. In Section \ref{sec:dis}, we discuss the implications of our Fe K$\alpha$ fits. Our results are summarized in Section \ref{sec:sum}.

\section{Observations}\label{sec:obs}

\begin{figure*}
\begin{center}
\begin{tabular}{c}
\resizebox{170mm}{!}{\includegraphics{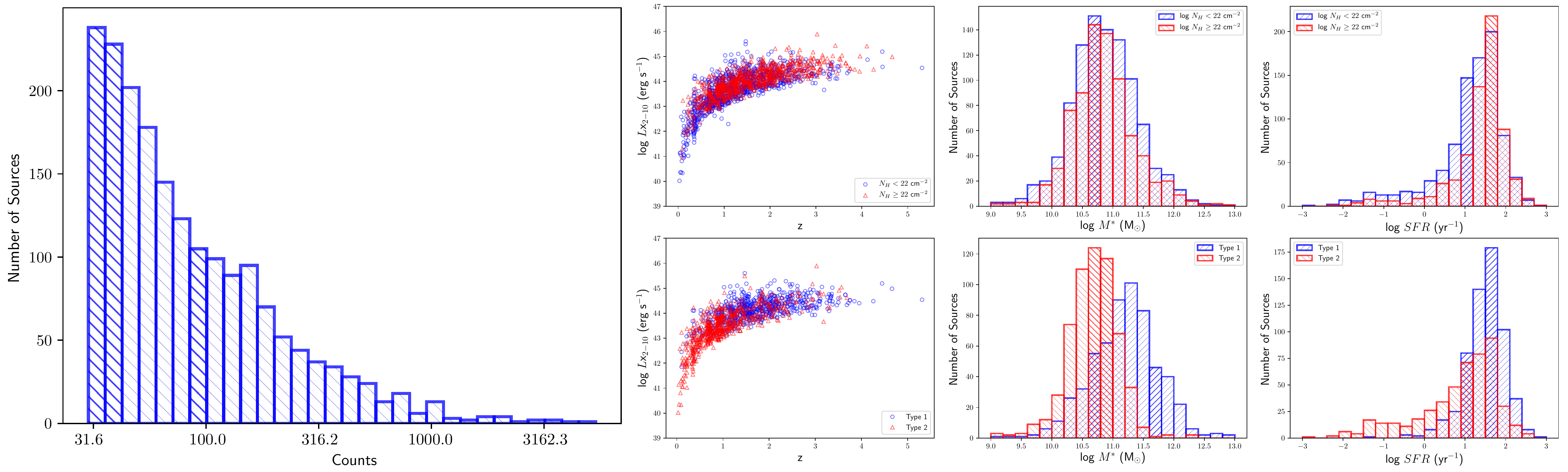}} \\
\end{tabular}
\caption{\textit{Chandra} COSMOS-Legacy survey selected sample properties: (top, left) distribution of source $0.5-7$ keV net counts. The remaining plots (columns from left to right) depict survey properties including: X-ray luminosity vs redshift, stellar mass, and star formation rate. These are then colored according to (top row) obscuring column density, and (bottom row) optically classified AGN type. \label{fig:img:cosmos}}
\end{center}
\end{figure*}

The 4.6Ms CCLS covers the 2 deg$^2$ COSMOS field to a depth of $2\times10^{-16}$ erg cm$^{-2}$ s$^{-1}$ in $0.5-2.0$ keV. The full catalog contains 4016 sources, with a full multiwavelength characterization and both photometric and spectroscopic redshifts available (\citealt{Mar16O}) and also stellar mass and SFR computed with spectral energy distribution fitting (\citealt{Suh19,Suh20}). Moreover, X-ray spectral analysis results are available for $\sim$1900 sources with more than 30 counts in the $0.5-7$ keV (\citealt{Mar16X}). We adopt the best-fit X-ray spectral models from this last catalog in our analysis, as described in Section \ref{sec:stack}.

\subsection{Source Selection}

From the full CCLS, we select sources with a minimum of 30 counts in $0.5-7.0$ keV, as in \citet{Mar16X}, where good fits to the continuum are available, to define our primary sample of $\sim1900$ sources.
The multiwavelength identifications made for the CCLS sources provide additional information permitting further division of the sample by redshift, as well as environmental properties (e.g., luminosity, obscuration, star formation rate). Distributions of these properties for the selected sample may be found in Figure \ref{fig:img:cosmos}, including $0.5-7.0$ keV net counts. 

\section{Stacking}\label{sec:stack}

\begin{figure}
\begin{center}
\resizebox{80mm}{!}{\includegraphics{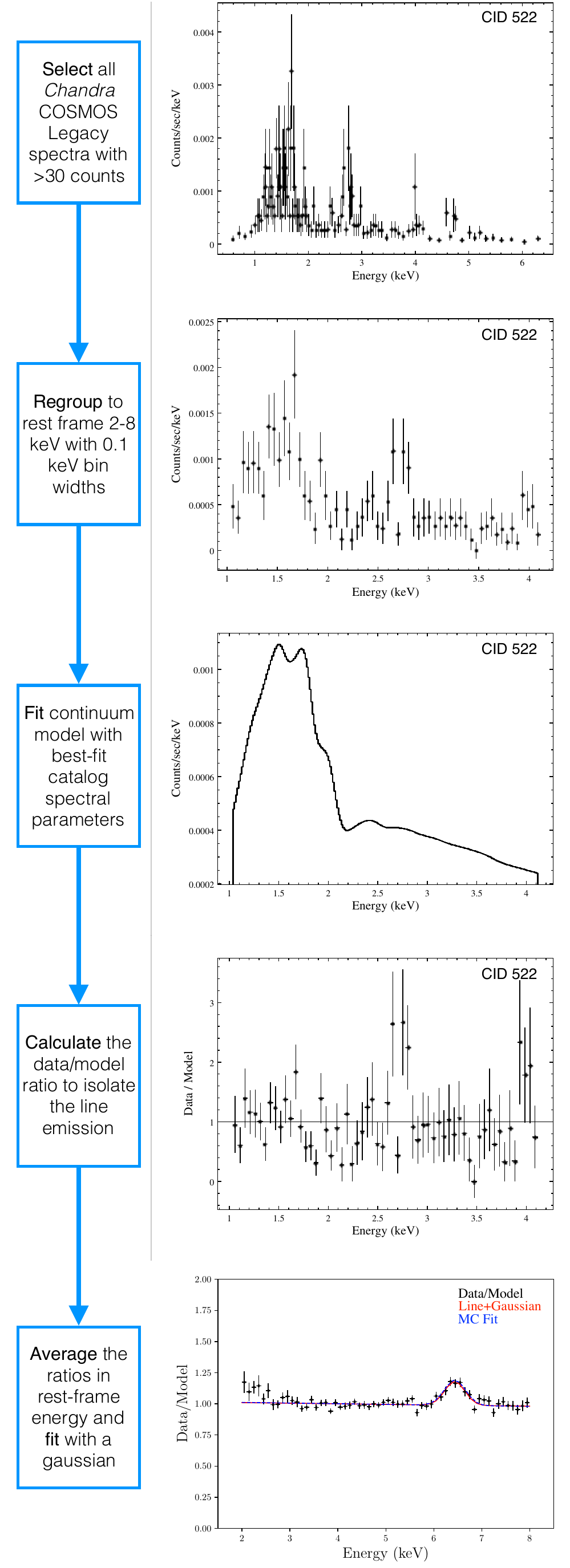}} \\
\caption{Schematic of the selected stacking method from reprocessed \textit{Chandra} spectra through Fe K$\alpha$ line emission Gaussian fits. The first four panels depict the stacking process for an example source, CID 522, while the last panel shows the averaged data/model ratio for all of the sources. \label{fig:method}}
\end{center}
\end{figure}

\begin{figure}
\begin{center}
\resizebox{90mm}{!}{\includegraphics{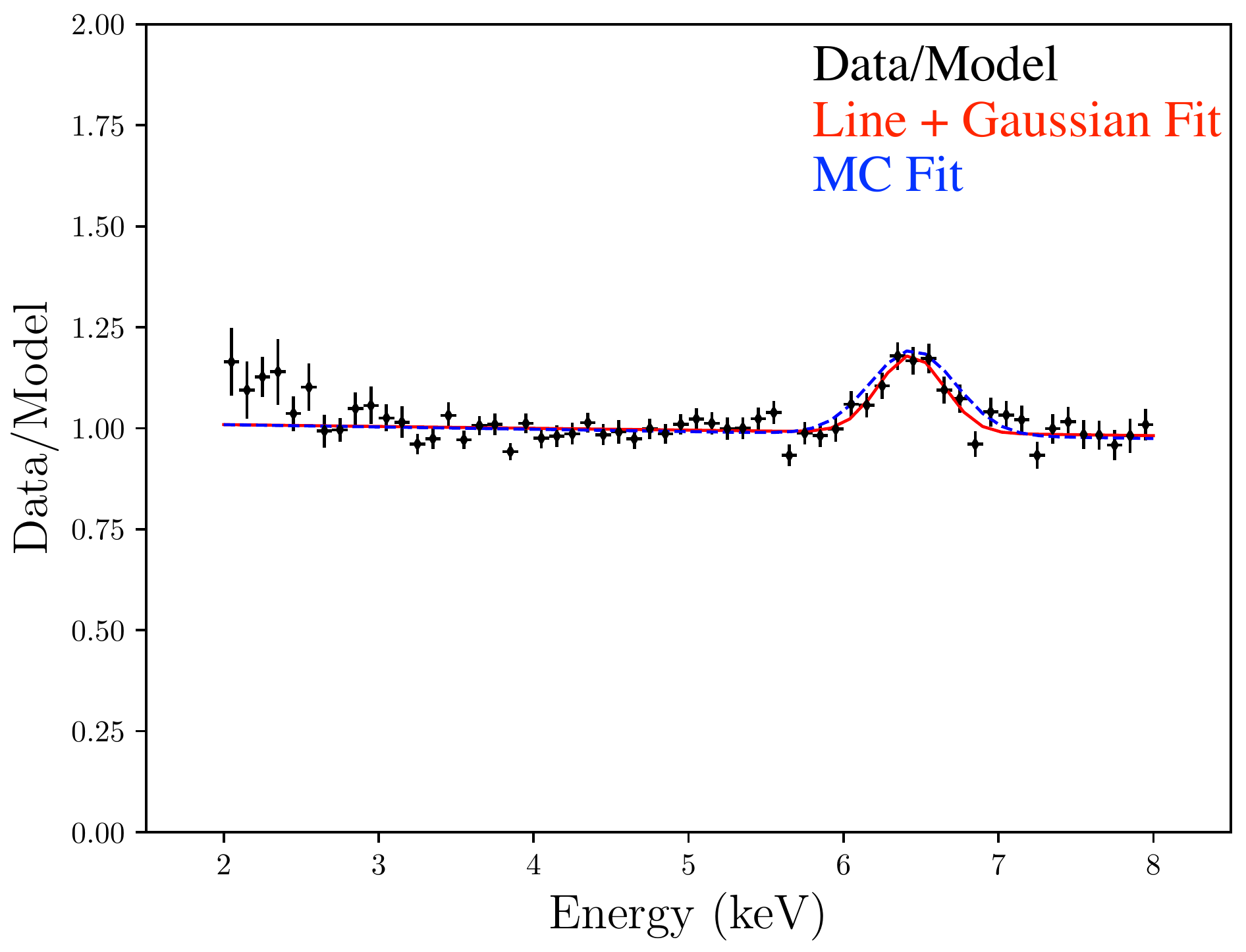}} \\
\caption{Average data/model ratio for the CCLS sources (black data points). This ratio is fit with a line + Gaussian model (red) to investigate the average properties of the Fe K$\alpha$ emission line at 6.4 keV. The model then undergoes a Monte Carlo analysis to determine the best-fit parameters (blue). \label{fig:fit:avg}}
\end{center}
\end{figure}

Our stacking analysis closely follows the methodology of \citealt{Cha12} (see also \citealt{Bru05,Cor08,Cor11,Wal15}). A schematic example of this method is shown in Figure \ref{fig:method}.
We primarily make use of CIAO 4.12\footnote{http://cxc.harvard.edu/ciao/}, CALDB 4.8.2\footnote{https://cxc.harvard.edu/caldb/}, and Sherpa\footnote{https://cxc.harvard.edu/sherpa/}, unless otherwise noted. 

For each source, we isolate an energy band corresponding to the \textit{Chandra} hard X-ray band, $2-8$ keV rest frame ($\frac{2.0}{1+z}-\frac{8.0}{1+z}$ keV), preferentially using spectroscopic redshifts, if available. The sources are regrouped using \texttt{grppha} in the HEASARC FTOOLS\footnote{http://heasarc.gsfc.nasa.gov/ftools} sub-package such that each channel is limited to this calculated waveband and rebinned in bin widths of 100 eV in rest frame ($\frac{0.1}{1+z}$ keV). 
With bin widths of 100 eV we capture important details of the spectra, especially surrounding the Fe K$\alpha$ emission line, while still maintaining significant counts within each energy bin, due to the spectral resolution of \textit{Chandra}.

The model continuum is defined based on the spectral analysis of the CCLS from \citet{Mar16X}. Our spectral fits use Cstat statistics (based on Cash statistics; \citealt{Cash79}) and proper modeling of the \textit{Chandra} background (\citealt{Mar16X}). Galactic absorption is fixed to the average column density observed in the direction of the COSMOS field ($\text{N}_H=2.6\times10^{20}$ cm$^{-2}$; \citealt{Kal05}). 

The source spectral models vary and are informed by the \citet{Mar16X} best fits. In \citet{Mar16X}, rather than assuming a single absorbed power law slope for every source, sources with $30<\text{counts}<70$ are fit with a fixed $\Gamma=1.9$, while in sources with higher counts the power law slopes are left free to vary. We adopt the best fit $\Gamma$ from this analysis for both of these count regimes. The photoelectric absorption intrinsic to the AGN is fixed directly from the best-fit column densities. An unabsorbed second power law component ($\Gamma_1=\Gamma_2$) is required for some sources in \citet{Mar16X} and has been included in our continuum fits for these individual sources. 
In some cases, the best-fit model parameters observed in \citet{Mar16X} are unphysical due to limited count statistics (e.g., $\Gamma$ is very flat). However, since we are only interested in an accurate phenomenological characterization of the continuum, unphysical continuum models do not adversely impact our analysis.

After fitting the continuum, we calculate the ratio of the data to our best-fit continuum model. 
Since our analysis first regroups each source into the same number of bins of equal width (100 eV rest frame) from the $2-8$ keV rest frame, the bins of each source spectra line up exactly in rest frame, which means that we can directly add and average all of the sources without further modifying the data/model ratios. We stack each source and calculate the average ratio for each bin in energy. This yields an average data/model ratio for the selected $\sim2000$ sources in the CCLS (Figure \ref{fig:fit:avg}). 

Since the focus of our analysis is on the Fe K$\alpha$ emission, our methodology deviates from that of \citet{Cha12}. Rather than multiplying our data/model ratios by a single assumed power law (E$^{-\Gamma}$) to convert it back to an averaged spectrum, we instead continue our analysis by fitting emission lines to the data/model ratio. We tested the accuracy of extracting the Gaussian widths using this more simplistic method on simulated data/model ratios, and find that while both methods are able to consistently recover the simulated Gaussian and equivalent widths to better than $0.1\%$ of the input width ($\sigma=0.1, 0.25, 0.35, 0.5$), fitting the data/model ratio is typically more precise by 6 decimal places.

The CCLS field is rich in multiwavelength data which uniquely allows us to break down the average data/model ratio as a function of observables, including: redshift, luminosity, star formation rate ($SFR$), stellar mass ($M^*$), optically-informed AGN type, obscuring column density ($N_H$), and observed $0.5-7.0$ keV spectral counts. Binning our sample by the total source spectral counts acts as a test to assess whether or not we are biasing our results with the highest S/N sources. Bins of AGN Type are separated into the Type 1/Type 2 designations as informed by their optical counterpart matches (see \citealt{Mar16O} for a description of the multiwavelength CCLS catalog match). Similarly, $N_H$ is separated into bins of unobscured ($\log N_H<20$ cm$^{-2}$; $20<\log N_H<22$ cm$^{-2}$) and obscured ($\log N_H>22$ cm$^{-2}$). 
While a majority of the ``unobscured'' sources are real unobscured sources, there may exist a population of heavily obscured sources ($\log N_H>23$ cm$^{-2}$) hiding within this subsample for which the standard spectral fitting in \citet{Mar16X} did not work properly. These candidate CT AGN are further analyzed in \citet{Lan18}.

For the remaining properties, we use dynamically sized bins where the bin size is determined by adding sources to achieve a minimum total number of counts ($0.5-7.0$ keV) in each bin. For this analysis, we use a minimum count threshold of $\sim40,000$ total counts. In this way, we are using the $0.5-7.0$ keV total counts as a proxy for achieving a high S/N in the average data/model ratio bins. We include the full, broad range of observed properties in the analysis, but note that only 21 sources fall below $\log M^*<9$, while only 7 sources fall above $z>4$.

\section{Iron K$\alpha$ Line Fitting}\label{sec:fit}

\subsection{Gaussian Line Fits}\label{ssec:gauss}
\begin{figure}
\begin{center}
\resizebox{78mm}{!}{\includegraphics{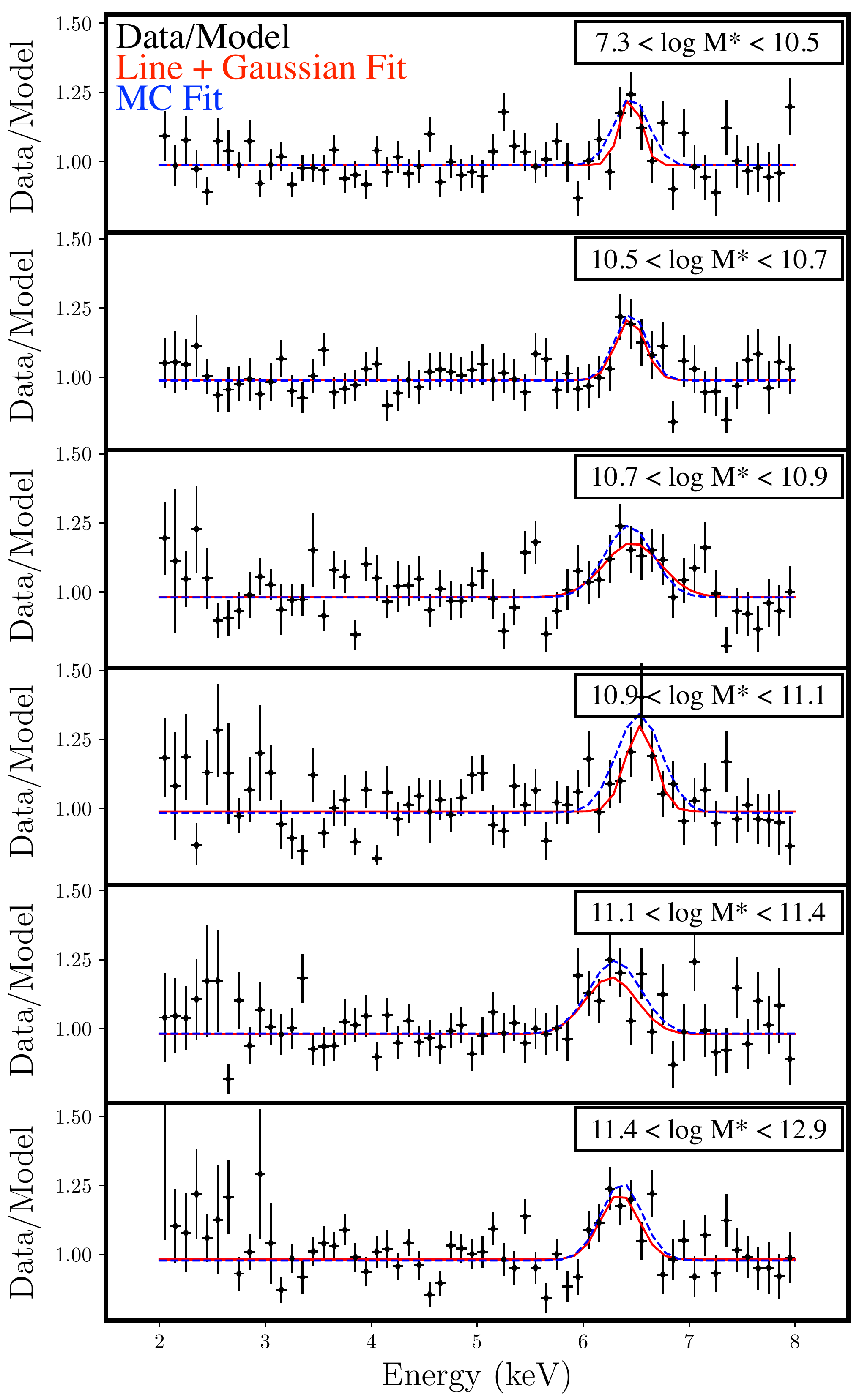}} \\
\caption{Data/Model ratio broken into bins of stellar mass (black) with the best line+Gaussian (red), and MC (blue) fits. \label{fig:fit:ms}}
\end{center}
\end{figure}

\begin{figure}
\begin{center}
\resizebox{76mm}{!}{\includegraphics{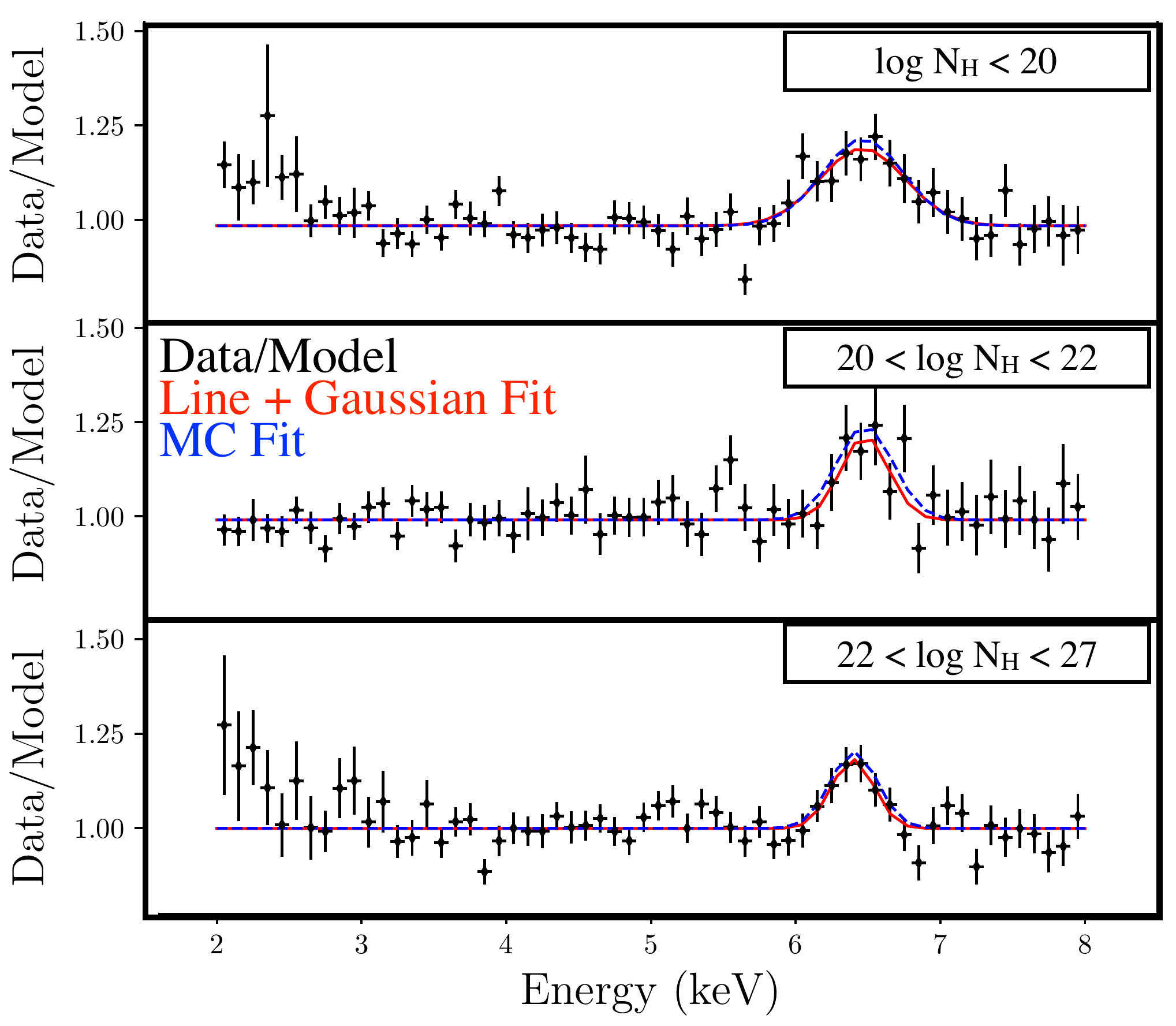}} \\
\caption{Data/Model ratio broken into bins of column density (black) with the best line+Gaussian (red), and MC (blue) fits. \label{fig:fit:nh}}
\end{center}
\end{figure}

By design we have removed the continuum in each spectrum by calculating the data/model ratio. Thus, for each source, the excess above unity consists of isolated X-ray emission lines that likely originate near the black hole. From these stacked average data/model ratios, we fit the excess above the continuum at 6.4 keV with a Gaussian Fe K$\alpha$ emission line model.

The fits are made using \texttt{curve\_fit}, part of the optimize module in the python scipy library. We assume that we have correctly redshifted all sources such that our data/model ratios are at rest frame ($z=0$) and the Fe K$\alpha$ emission line is represented by a Gaussian frozen at 6.4 keV. The function we use is an additive model of a straight line continuum (representing where the model exactly fits the data, i.e., unity) and Gaussian shape: $ratio = (m\times E + b) + C \times \exp(-(E-6.4)^{2}/2\sigma_g^2)$, where $m$ is the line slope, $b$ is the line normalization, $C$ is the Gaussian amplitude, and $\sigma_g$ is the Gaussian width. Our initial conditions and [boundary conditions] are: $m=0$ $[-0.1\,,\,0.1]$, $b=1.0$ $[0.9\,,\,1.1]$, $C=1.0$ $[0.0\,,\,2.0]$, and $\sigma_g=0.1$ $[0.1\,,\,0.5]$. We used a Monte Carlo method to randomly resample the data/model ratios ($N=100$) and determine the best fit. We fit the line+Gaussian shape for the average stacked spectrum (Figure \ref{fig:fit:avg}), as well as for each binned parameter (e.g., $M^*$, Figure \ref{fig:fit:ms}; $N_H$, Figure \ref{fig:fit:nh}; Additional parameter fits included in Appendix \ref{app:fit}).

To test our assumption that the average Fe K$\alpha$ emission line is centered at 6.4 keV and therefore not significantly impacted by e.g., different ionization states, or residual redshift effects, we allowed the centroid of the Gaussian line to vary, finding a best-fit centroid line energy of $6.45$ keV. We find that our results are not significantly impacted by setting the Gaussian line energy as a free parameter (to within a few percent), and have thus adopted a fixed line energy for simplicity and to limit possible degeneracies in the fit (Table \ref{tab:fits}).
We also test the impact of using photometric redshifts to fit the CCLS sample, when spec-z are not available, as is the case for $\sim$31\% of the selected sample. We find the Gaussian widths fit using only sources with spec-z are within the errors of the full sample.

\subsection{Relativistic Line Fits}\label{ssec:rel}

The Gaussian fit is more simplistic than the relativistic line models that are typically used to measure $a$, the spin measurement parameter ($a=cJ/GM^2$, where M is the black hole mass, J is the angular momentum; $-1<a<1$) for sources with hundreds of thousands of counts. 
It is important to note that while we cannot duplicate the detail of an in-depth spin analysis of an individual, bright AGN using a stacking method that itself may smooth over interesting physics and could introduce systematic errors, our intent is to determine the ``average'' black hole spin and compare this spin with e.g., redshift, and other environmental variables.
While we must make some assumptions to test these more complex relativistic fits (e.g., frozen model parameters), what we gain are new insights into the evolution of SMBHs and their host galaxies that are currently limited by observational constraints.

\subsubsection{\texttt{diskline}}

\begin{figure}
\begin{center}
\resizebox{80mm}{!}{\includegraphics{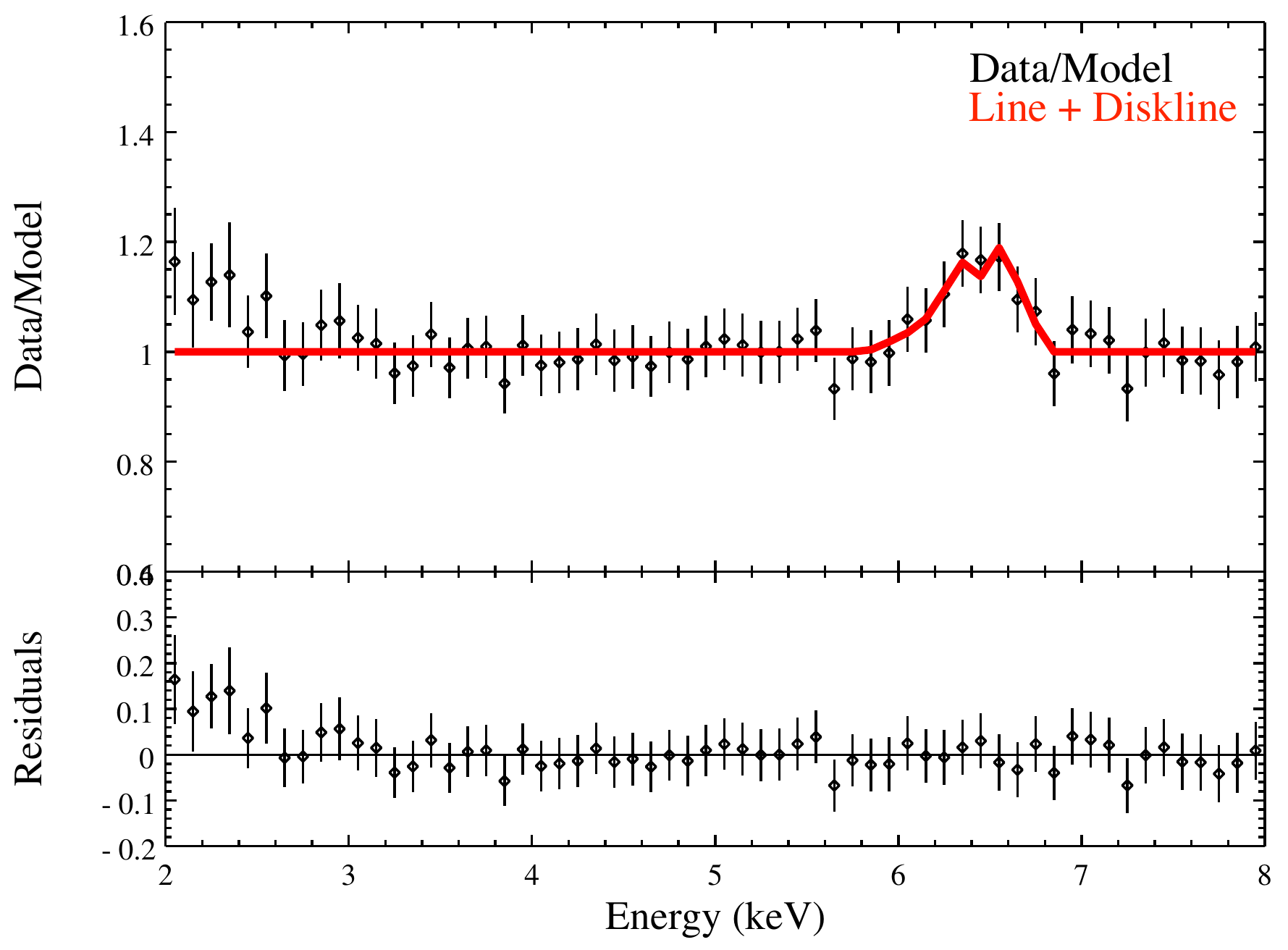}} \\
\caption{Average data/model ratio for the CCLS sources (black data points). This ratio is fit with a line + \texttt{diskline} model (red) with black hole spin parameter fixed at $a=0$. \label{fig:fit:dl}}
\end{center}
\end{figure}

We attempt to constrain the black hole spin parameter starting with the simplest of these relativistic line models, \texttt{diskline}, representing broadened line emission from the inner accretion disk with a non-spinning black hole ($a=0$; \citealt{Fab89}). We first freeze all parameters, except the normalization, at their default values (including LineE = 6.4 keV). We then systematically thaw these model parameters and fit the data using the Nelder-Mead Simplex optimization method; we test the goodness of fit with $\chi^2$ minimization (details of these fits are included in Appendix \ref{app:rel}). Ultimately we find a very good fit ($r\chi^2 = 0.413$) using this relativistic line model that assumes no black hole spin (Figure \ref{fig:fit:dl}).

\subsubsection{\texttt{relline}}

\begin{figure}
\begin{center}
\resizebox{80mm}{!}{\includegraphics{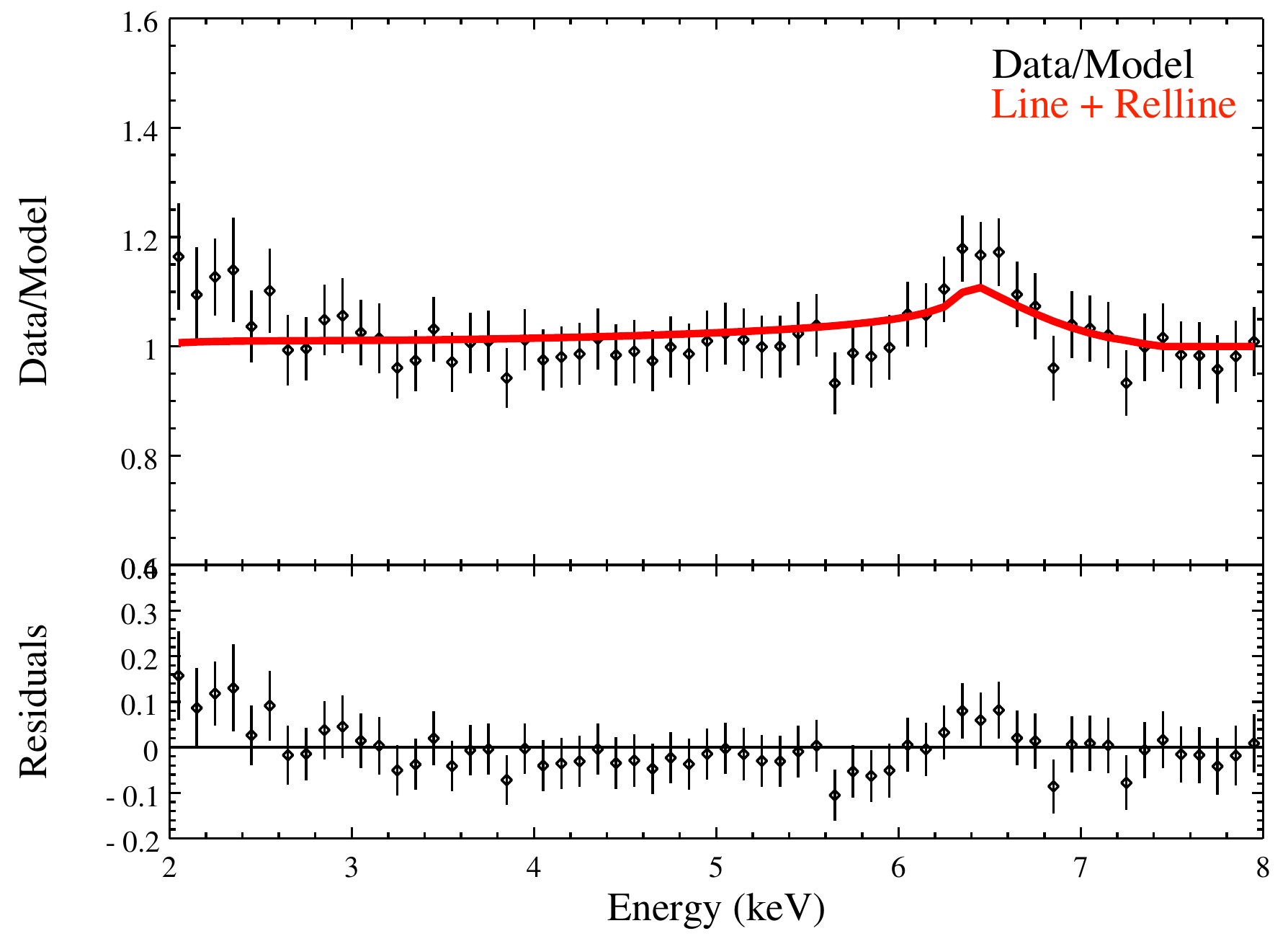}} \\
\caption{Average data/model ratio for the CCLS sources (black data points). This ratio is fit with a line + \texttt{relline} model (red) with the black hole spin parameter, normalization, and inclination left to vary. \label{fig:fit:rl}}
\end{center}
\end{figure}

We find that while \texttt{diskline} provides a good fit to the average data/model ratio, it is not a fully relativistic model and does not adequately parameterize light bending. Thus, rather than assuming a static value for the black hole spin, we instead incorporate a more physical relativistic line model, \texttt{relline}, from \texttt{relxill} v1.3.3 (\citealt{Dau10,Dau13,Dau14}) to fit the black hole spin parameter, $a$. 
As with \texttt{diskline}, we start with the fewest thawed parameters before increasing the fit complexity by introducing additional free parameters. For all fits, we keep $a$ and the normalization of this relativistic line free. The other parameters of interest are kept static, including Index1$=$6, Index2$=$3, and Rbr$=$6, where the emissivity for the corona is defined as Index1, and Index2, for the regions Rin to Rbr, and Rbr to Rout, respectively. We first investigate how the inclination of the accretion disk impacts the shape of the relativistic line (Appendix, Figure \ref{fig:fit:rlincl}) for inclination values of 45, 30, and 60 degrees, before testing the parameter space surrounding Index1 and Rbr (Appendix, Figure \ref{fig:fit:rlparam}). The best-fit relativistic line model left inclination free to vary and set Index1$=10$, and found $a=0.76\pm0.02$ ($r\chi^2=0.598$). The Fe K$\alpha$ emission line is not completely fit in this model, as shown in Figure \ref{fig:fit:rl}, however, and a more complex model may be needed.

\subsubsection{\texttt{relline} + Gaussian}

\begin{figure}
\begin{center}
\resizebox{80mm}{!}{\includegraphics{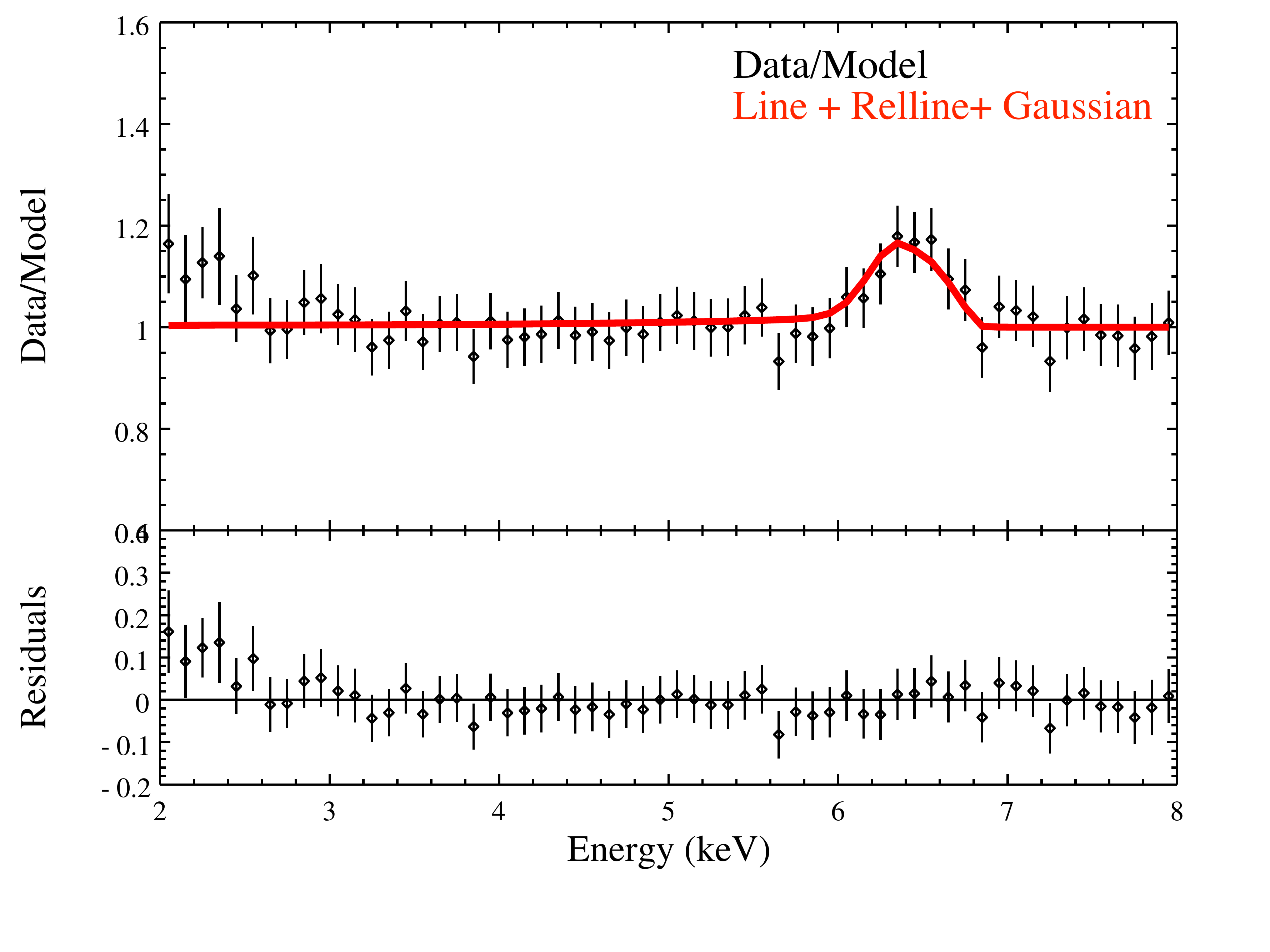}} \\
\caption{Average data/model ratio for the CCLS sources (black data points). This ratio is fit with a line + \texttt{relline} + Gaussian model (red) with the black hole spin parameter, normalization, line energy, and inclination left to vary. The Gaussian width and normalization are also left free to vary in this best-fit model. \label{fig:fit:rlg}}
\end{center}
\end{figure}

Since the single \texttt{relline} model left residuals around the Fe K$\alpha$ 6.4 keV emission line, we added a narrow Gaussian at 6.4 keV to the model to represent the combination of relativistically broadened and narrow Fe emission (a typical feature observed in AGN; \citealt{Yaq04}).
A two component (broad + narrow) model fit to the Fe K$\alpha$ emission is also consistent with the best-fit model found for the stacked AGN in the \textit{2XMM} catalog (\citealt{Cha12}), and for the stacked 263 \textit{XMM}-selected unabsorbed AGN spectra from the V\`eron-Cetty and V\`eron catalogs (\citealt{Fal14}).

Similar to the process for the single \texttt{relline} model, we systematically stepped through the parameter space, starting with inclination of the \texttt{relline} accretion disk (Appendix, Figure \ref{fig:fit:rgincl}). The additional Gaussian width was frozen at 0.1 keV, while the normalization was left free. We then tested the parameter space surrounding Index1 and Rbr (Appendix, Figure \ref{fig:fit:rgparam}), and similarly found a better fit when Index1$=10$. Finally, we varied the width of the additional Gaussian in addition to the \texttt{relline} parameters (Appendix, Figure \ref{fig:fit:rgbest}). The best-fit relativistic line + Gaussian model set the \texttt{relline} inclination and line energy, and the Gaussian width free to vary (Figure \ref{fig:fit:rlg}), although many of the model combinations we tried yielded very good fits. In this model, Index1$=10$, Index2$=3$, and Rbr$=10$, and the black hole spin parameter was found to be $a=0.62~\substack{+0.07 \\-0.17}$ ($r\chi^2=0.456$).

\section{Discussion}\label{sec:dis}

\begin{figure*}
\begin{center}\footnotesize
\begin{tabular}{cc}
\resizebox{70mm}{!}{\includegraphics{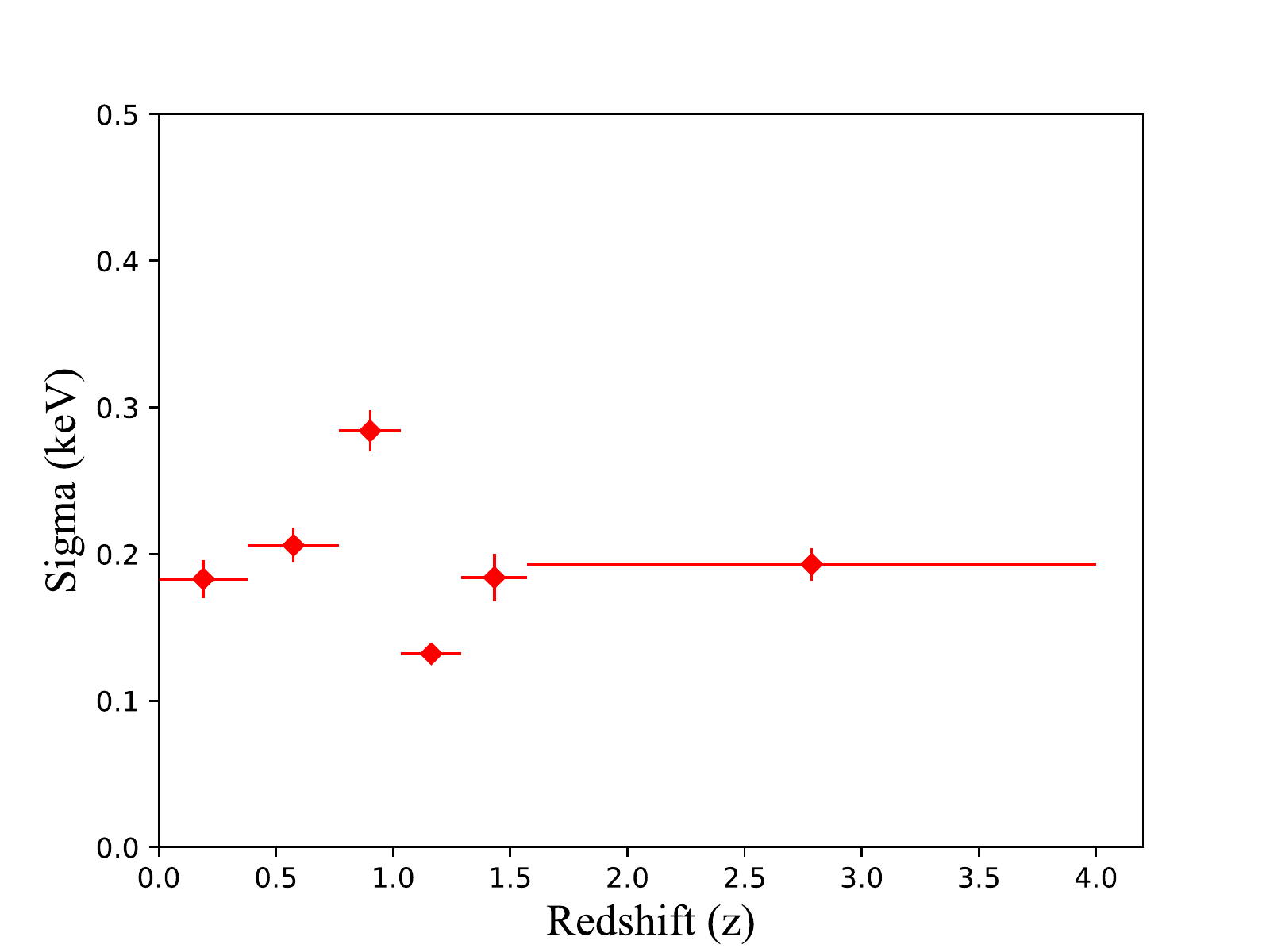}} &  \resizebox{70mm}{!}{\includegraphics{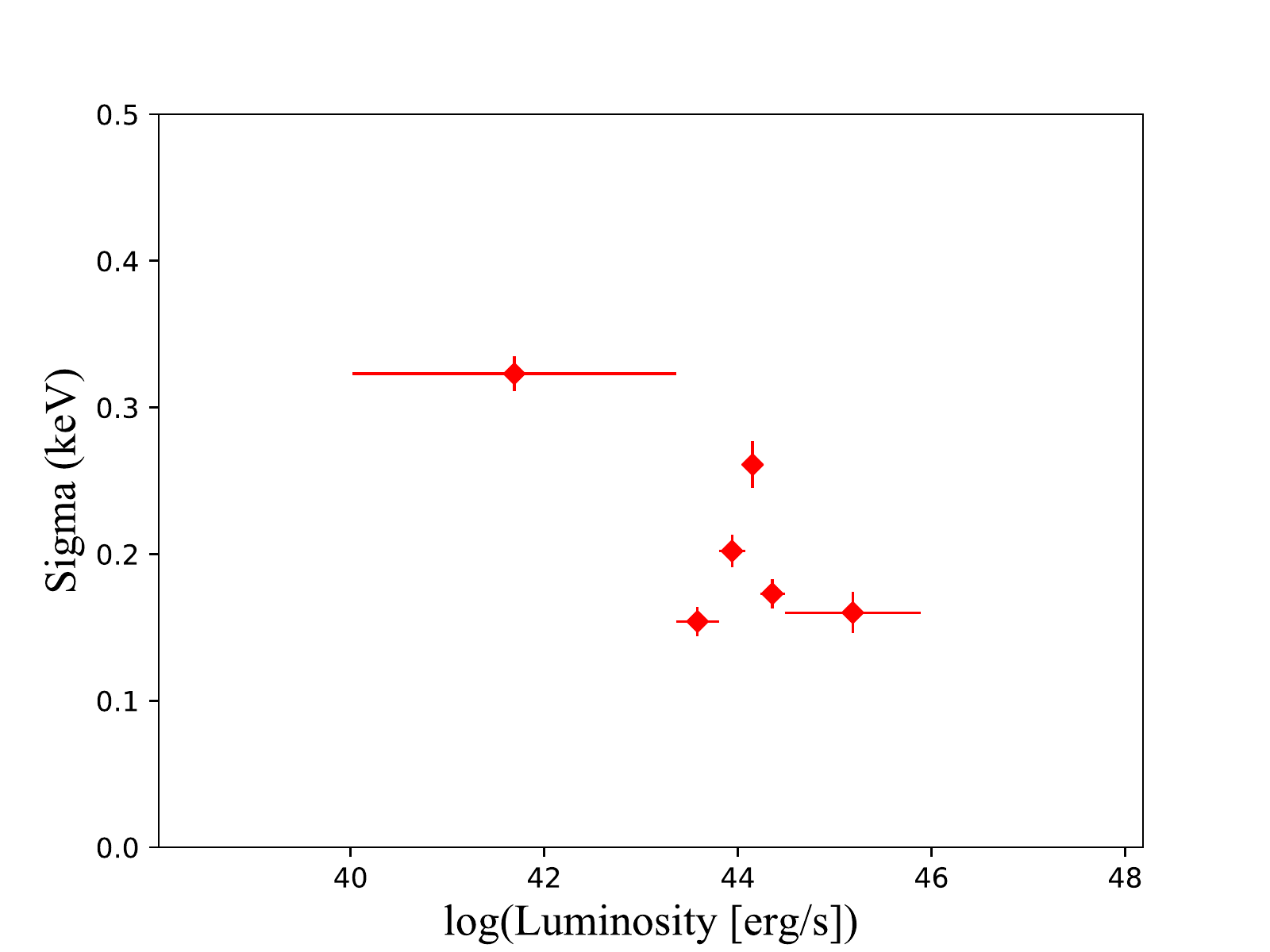}} \\
\resizebox{70mm}{!}{\includegraphics{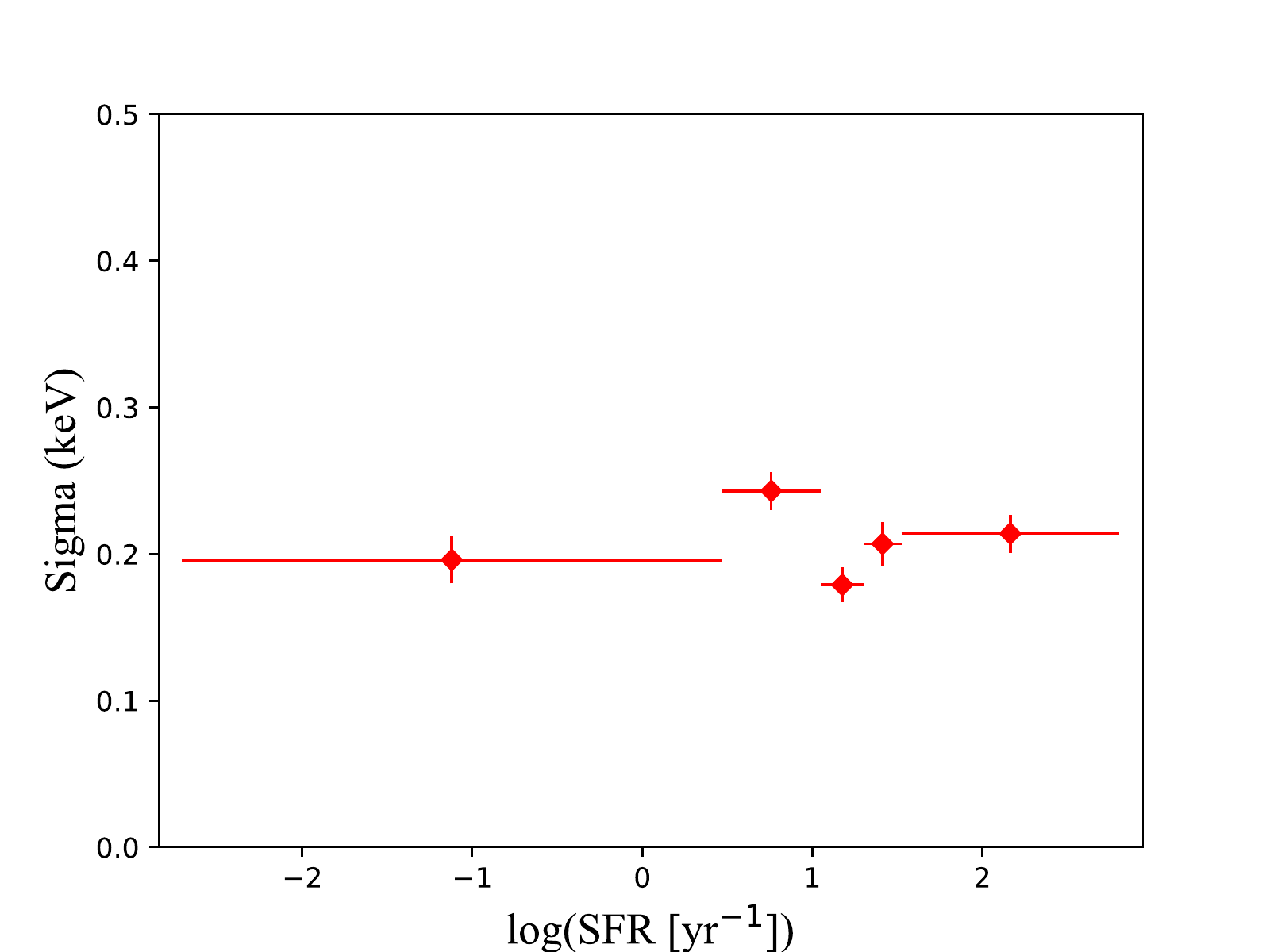}} &  \resizebox{70mm}{!}{\includegraphics{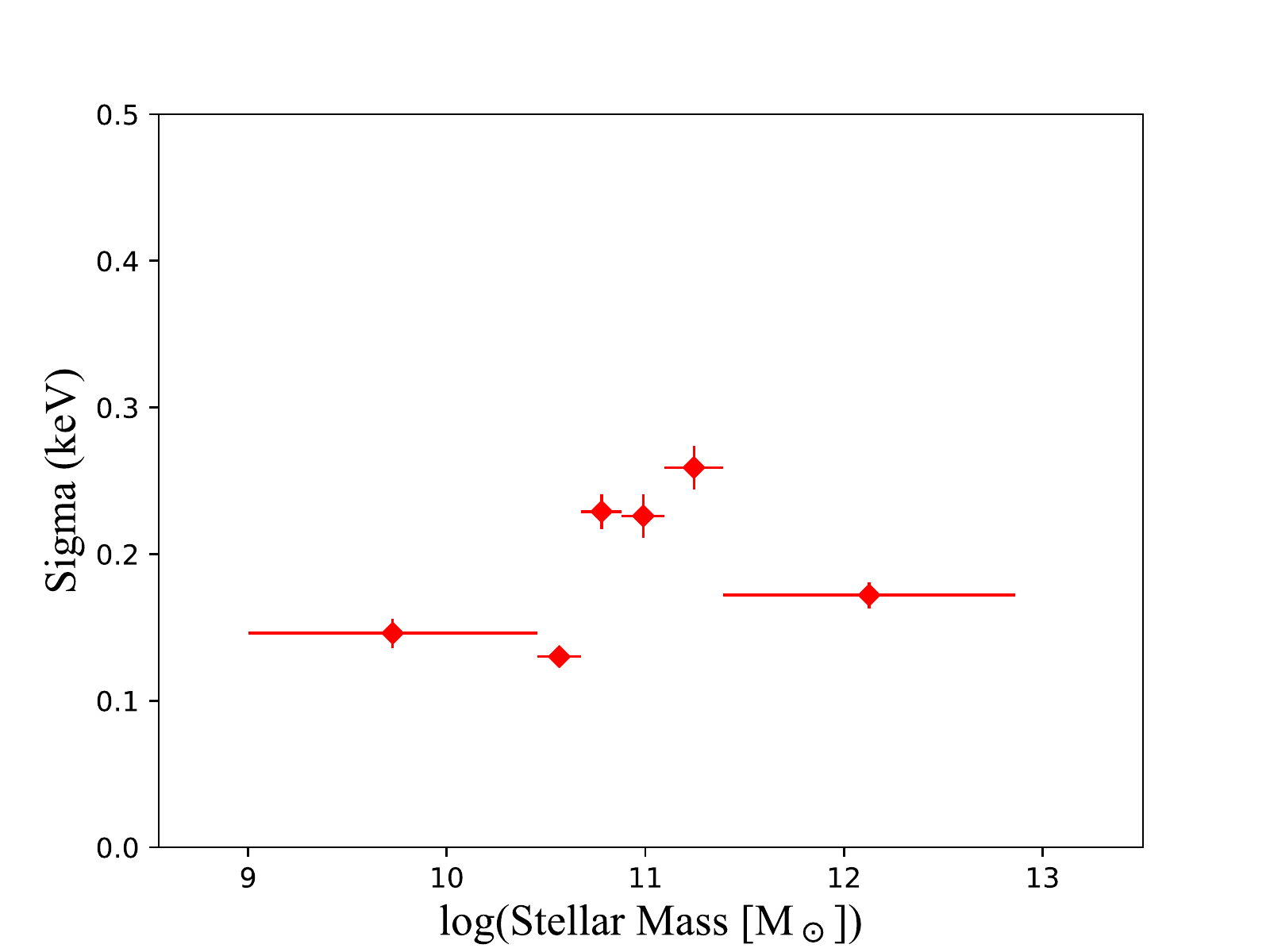}} \\
\resizebox{70mm}{!}{\includegraphics{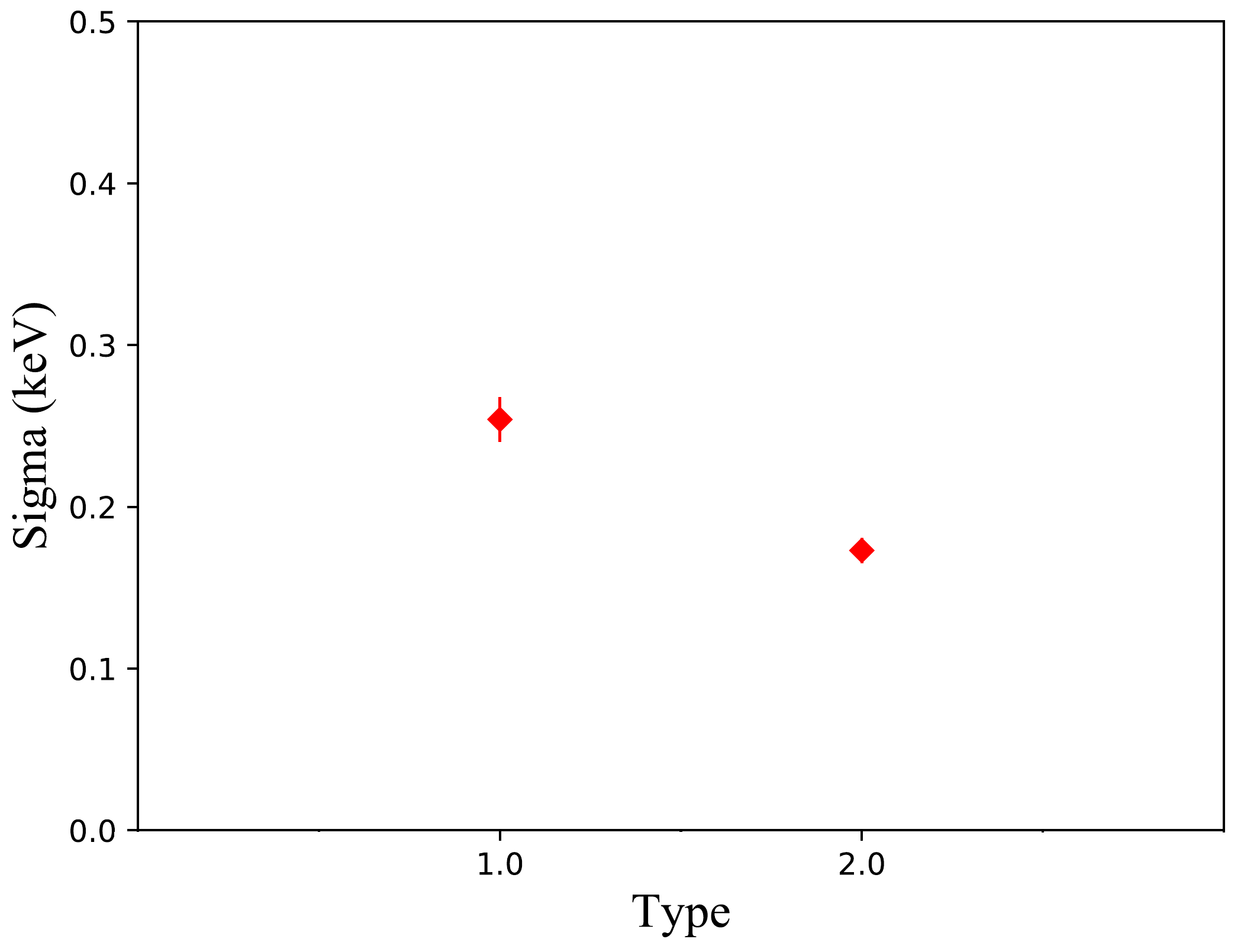}} &  \resizebox{70mm}{!}{\includegraphics{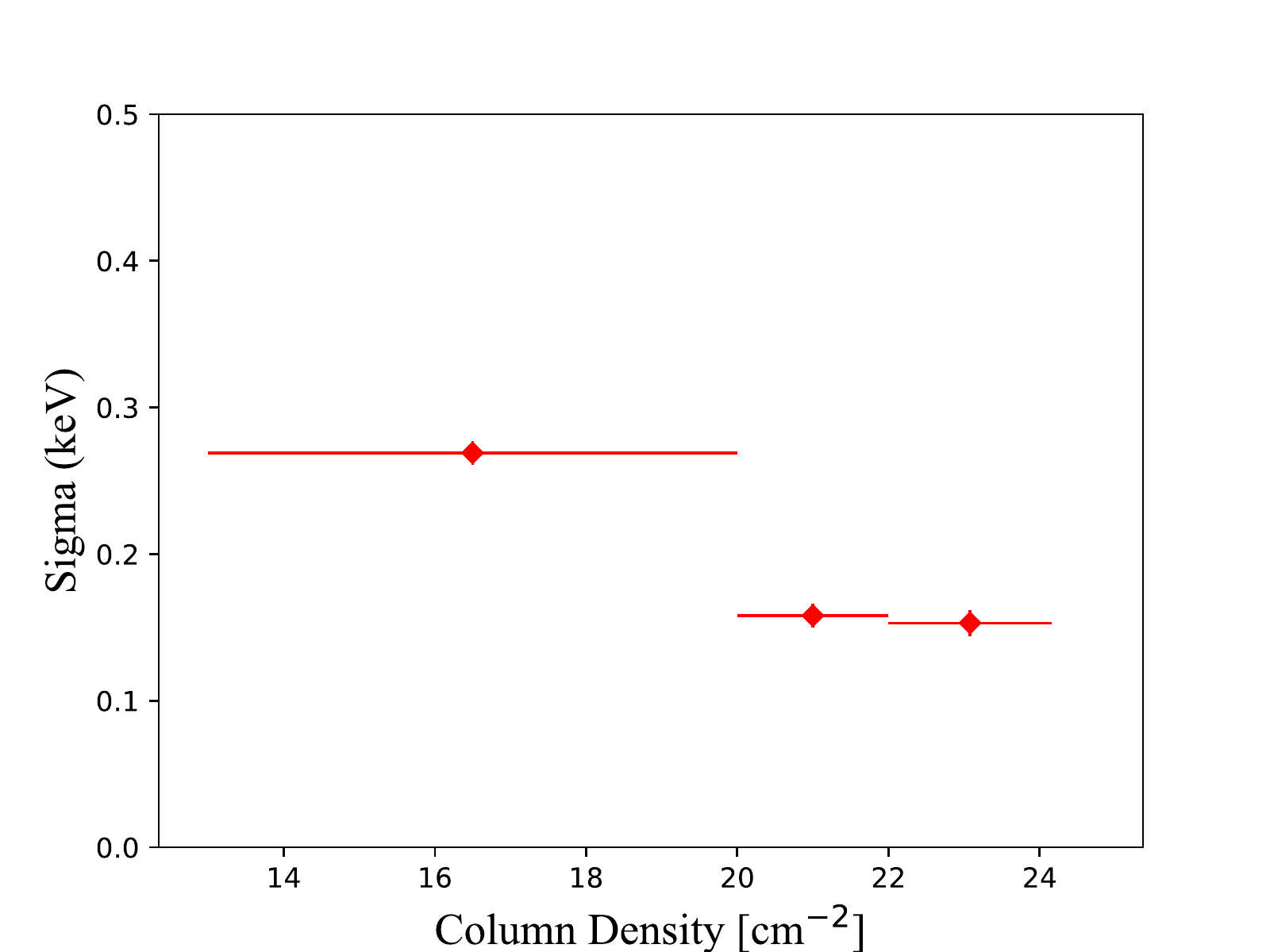}} \\
 \multicolumn{2}{c}{\resizebox{70mm}{!}{\includegraphics{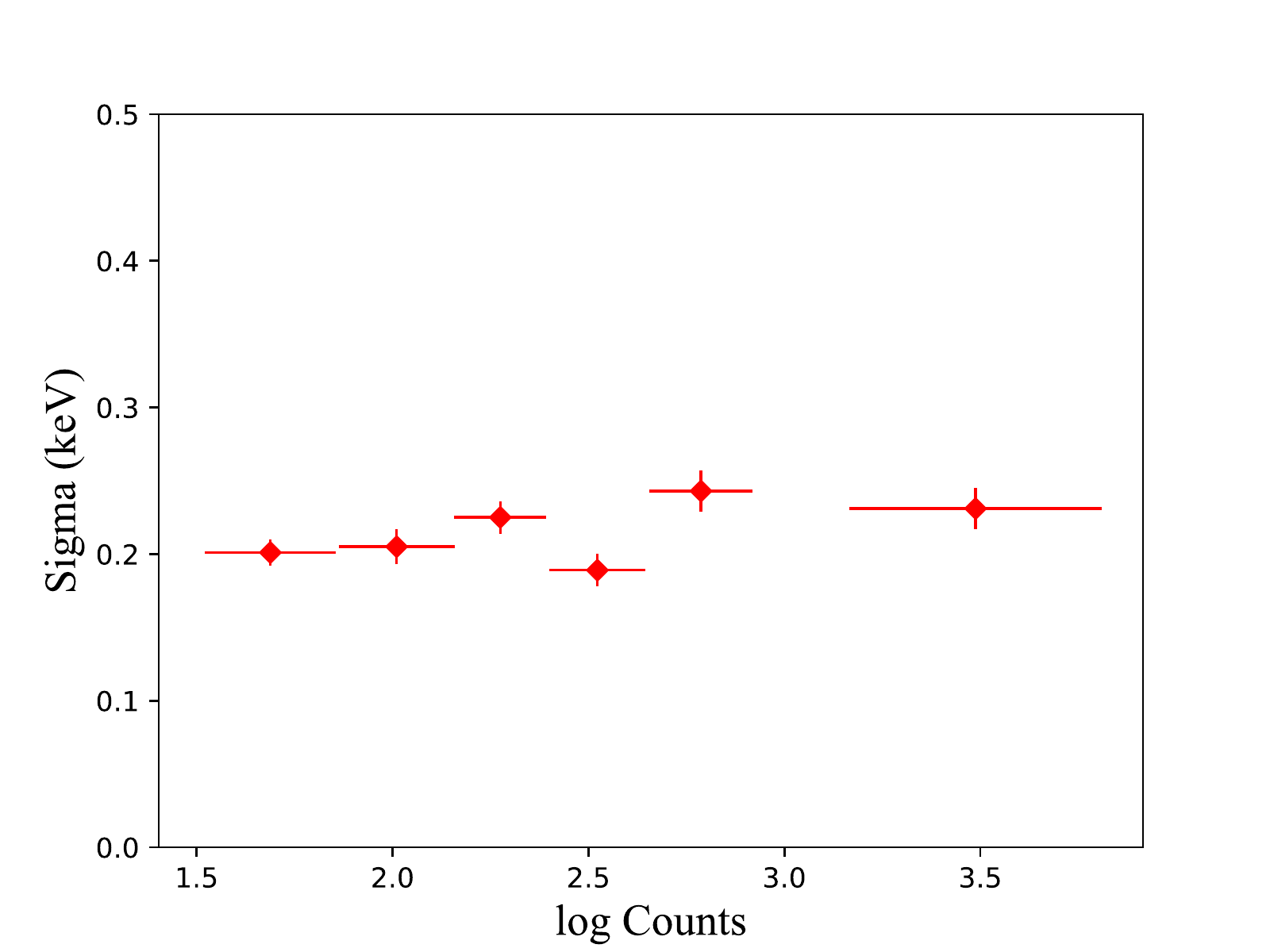}}} \\
\end{tabular}
\caption{Evolution of the best-fit Gaussian widths in bins of observational properties. \label{fig:fit:gauss}}
\end{center}
\end{figure*}

\begin{figure}
\begin{center}
\resizebox{75mm}{!}{\includegraphics{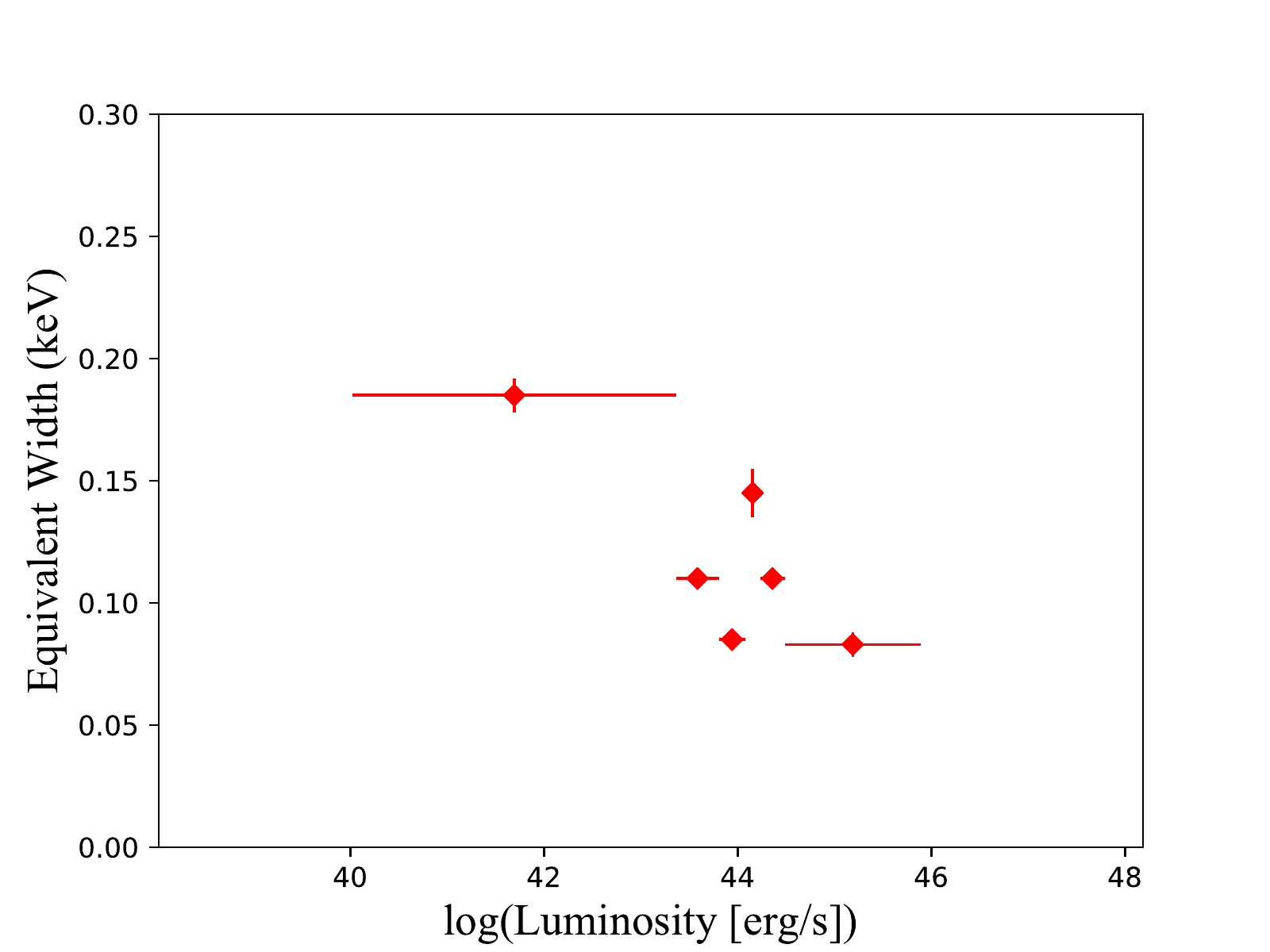}} \\
\caption{Evolution of the equivalent widths calculated from the best-fit Gaussian models for the CCLS sample in bins of $2-10$ keV X-ray luminosity. \label{fig:ew}}
\end{center}
\end{figure}

\begin{figure*}
\begin{center}\footnotesize
\begin{tabular}{cc}
\resizebox{70mm}{!}{\includegraphics{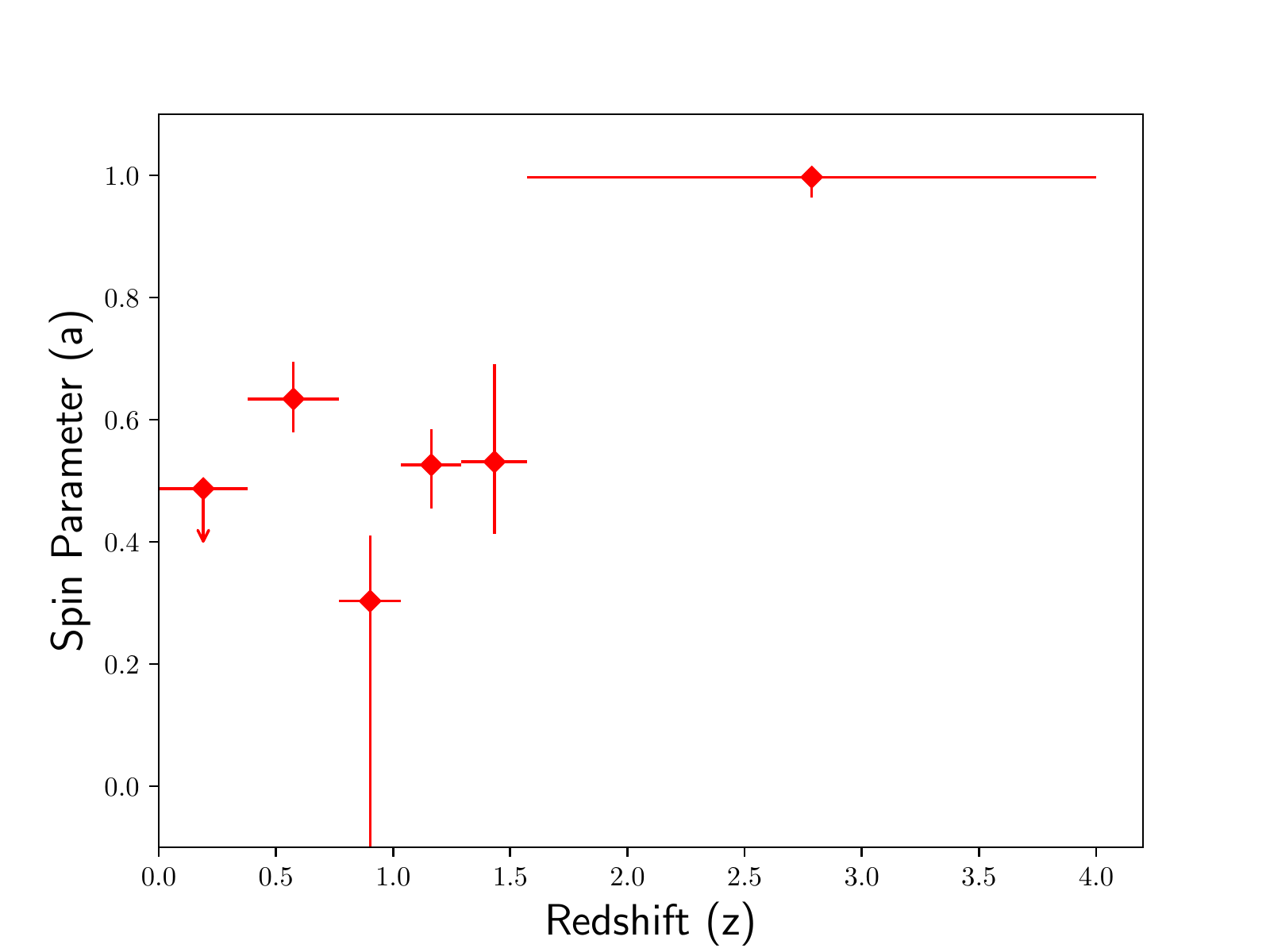}} &  \resizebox{70mm}{!}{\includegraphics{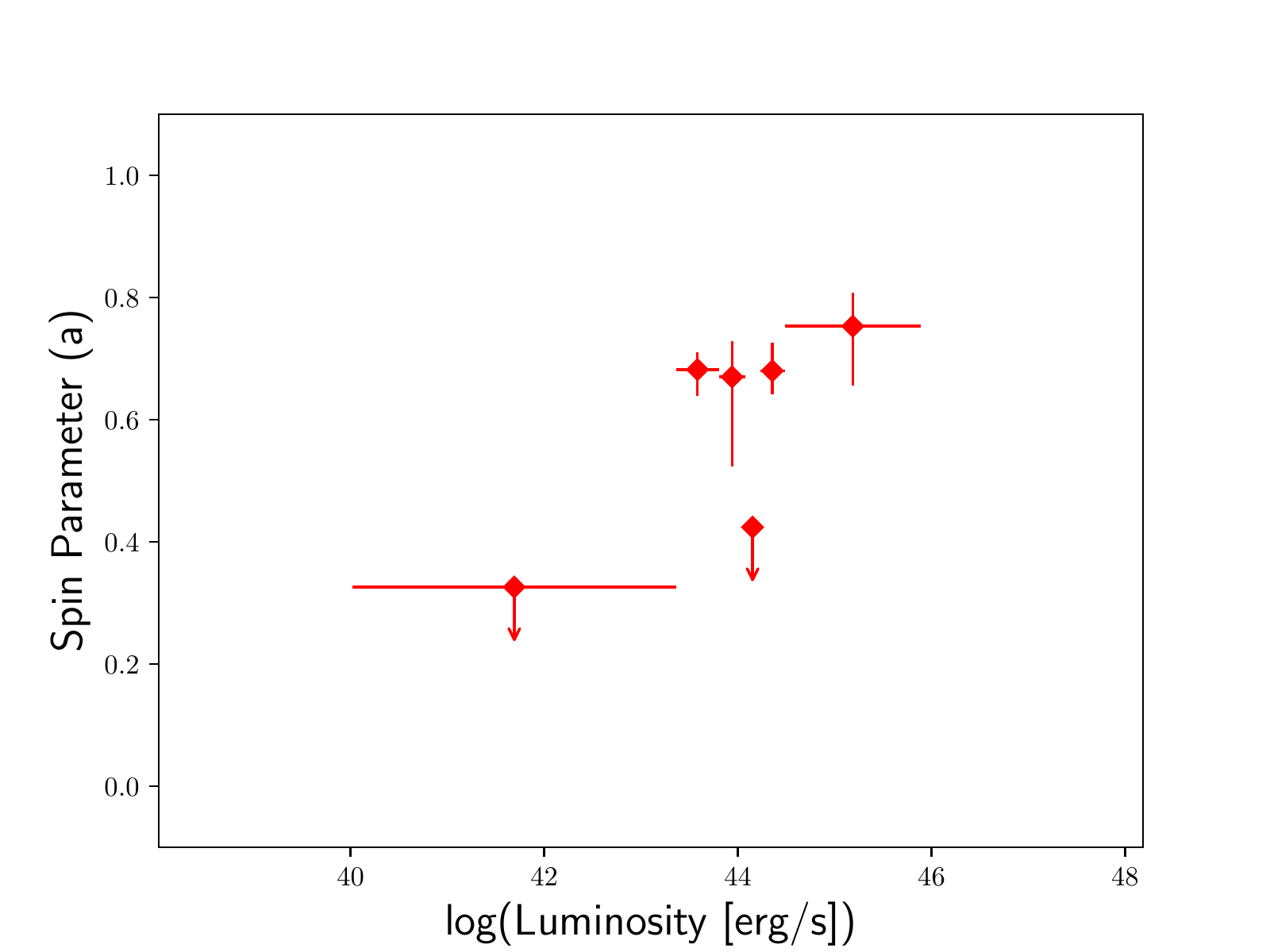}} \\
\resizebox{70mm}{!}{\includegraphics{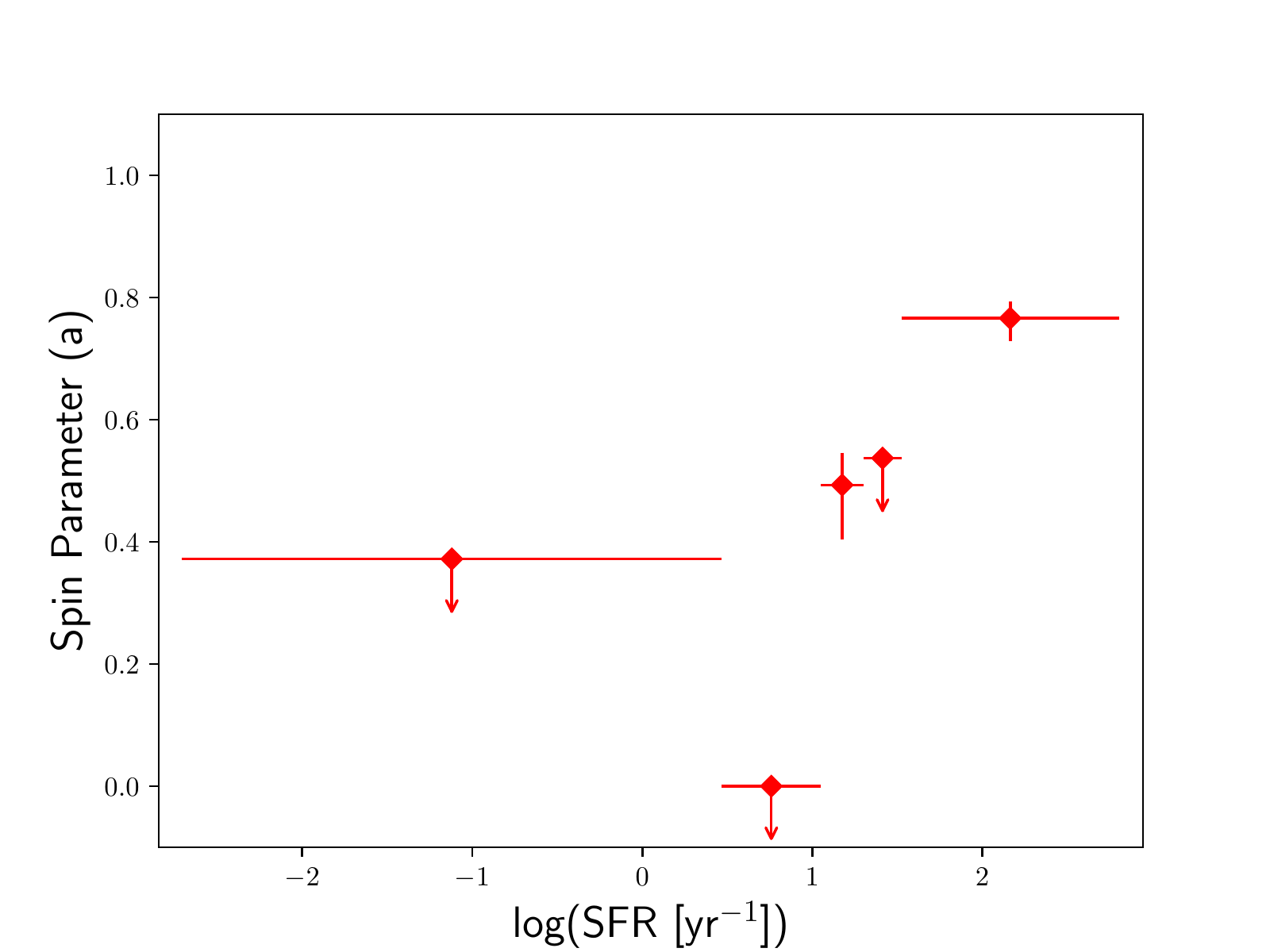}} &  \resizebox{70mm}{!}{\includegraphics{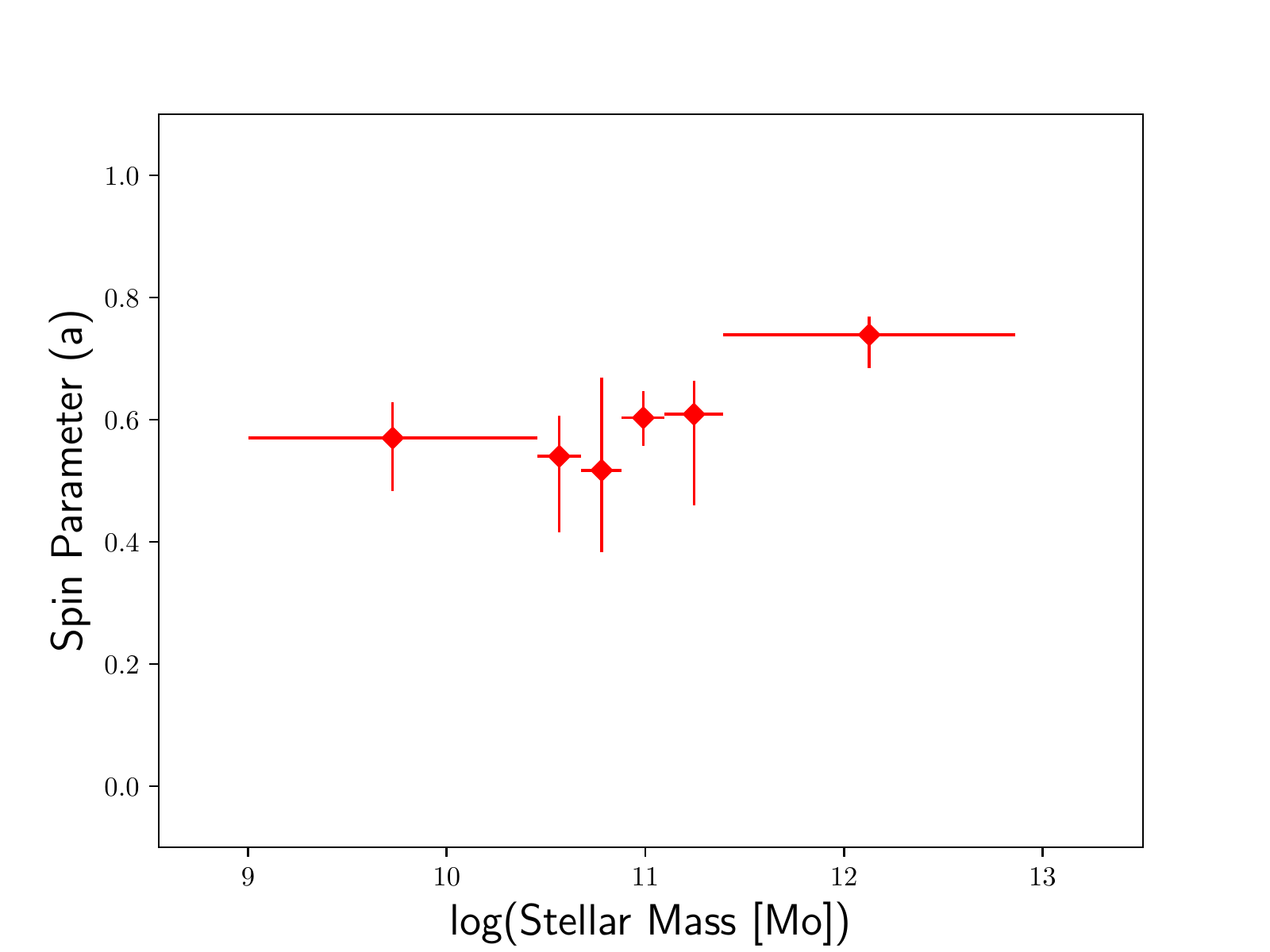}} \\
\resizebox{70mm}{!}{\includegraphics{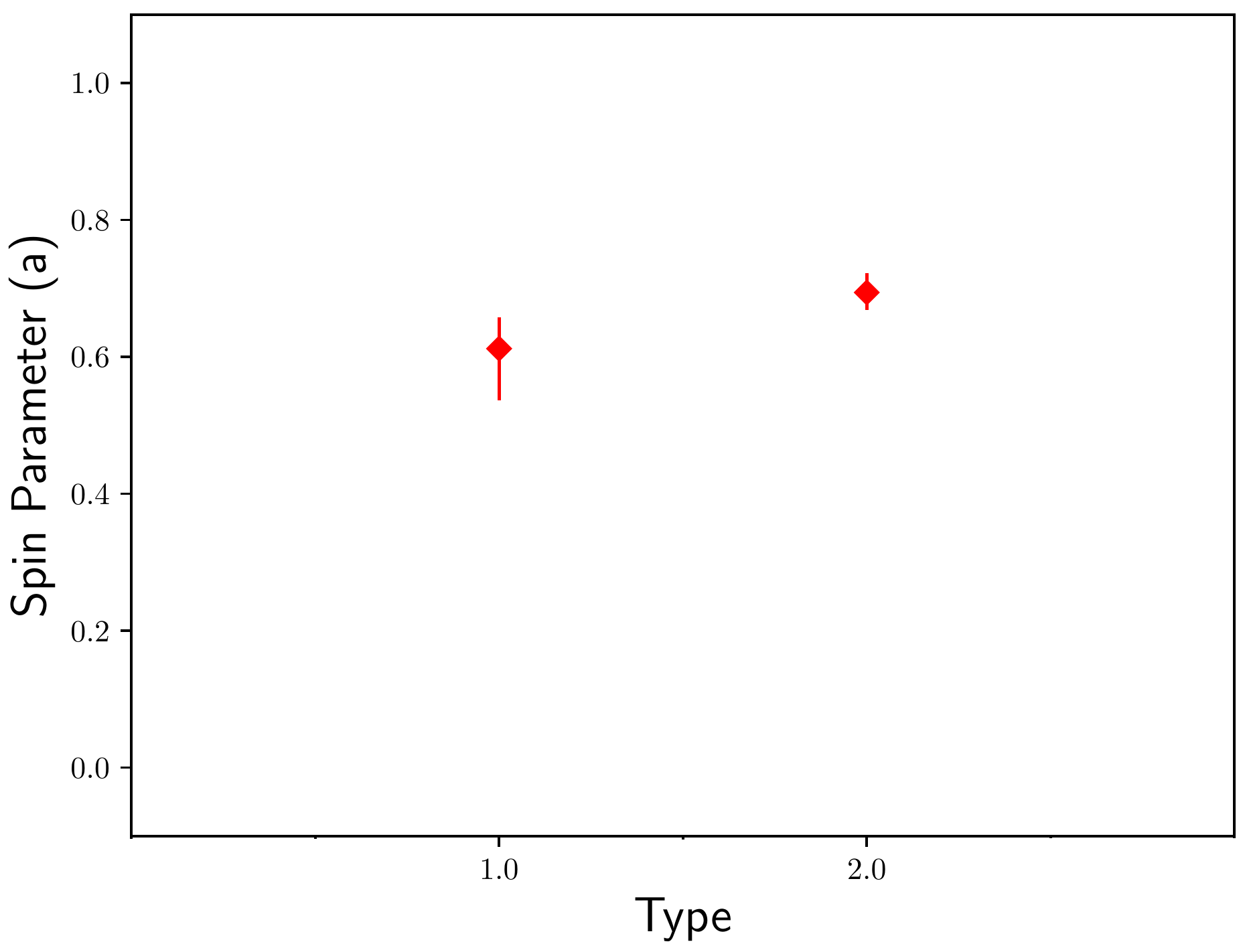}} &  \resizebox{70mm}{!}{\includegraphics{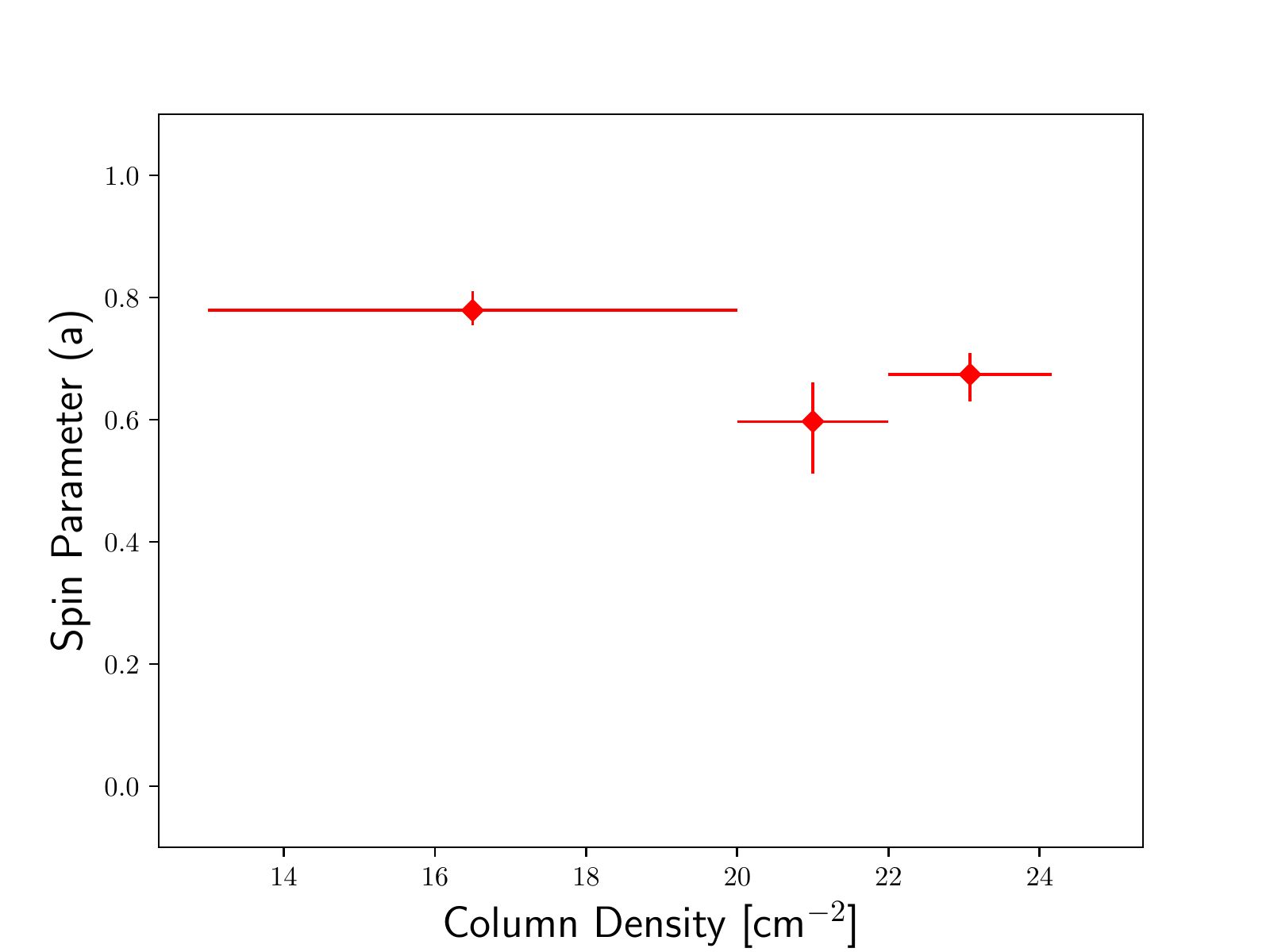}} \\
 \multicolumn{2}{c}{\resizebox{70mm}{!}{\includegraphics{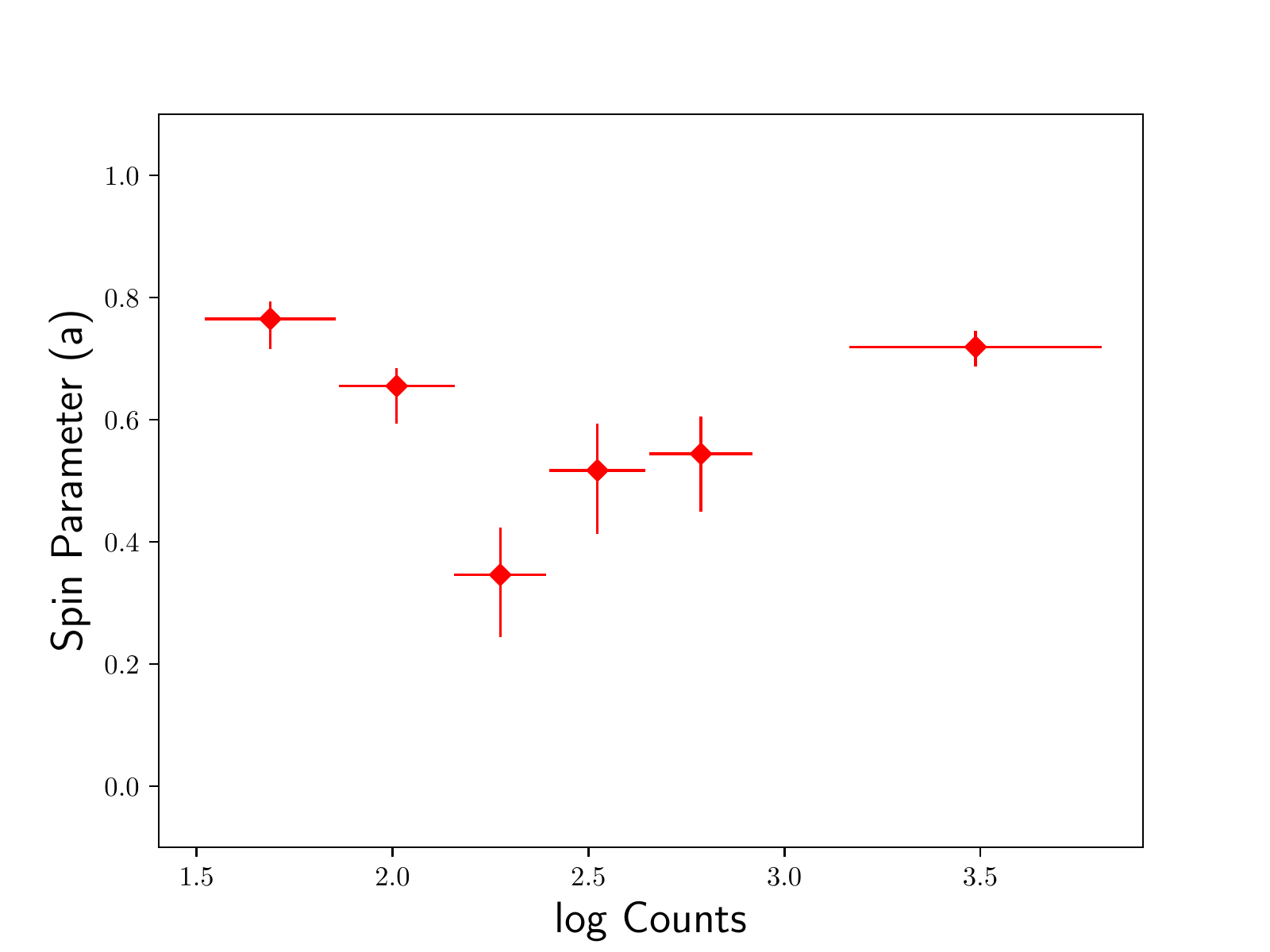}}} \\
\end{tabular}
\caption{Evolution of the best-fit \texttt{relline} spin parameter \textit{a} in bins of observational properties. Arrows indicate where the spin parameter represents an upper limit and is not well constrained. \label{fig:fit:relline}}
\end{center}
\end{figure*}

\begin{figure*}
\begin{center}\footnotesize
\begin{tabular}{cc}
\resizebox{70mm}{!}{\includegraphics{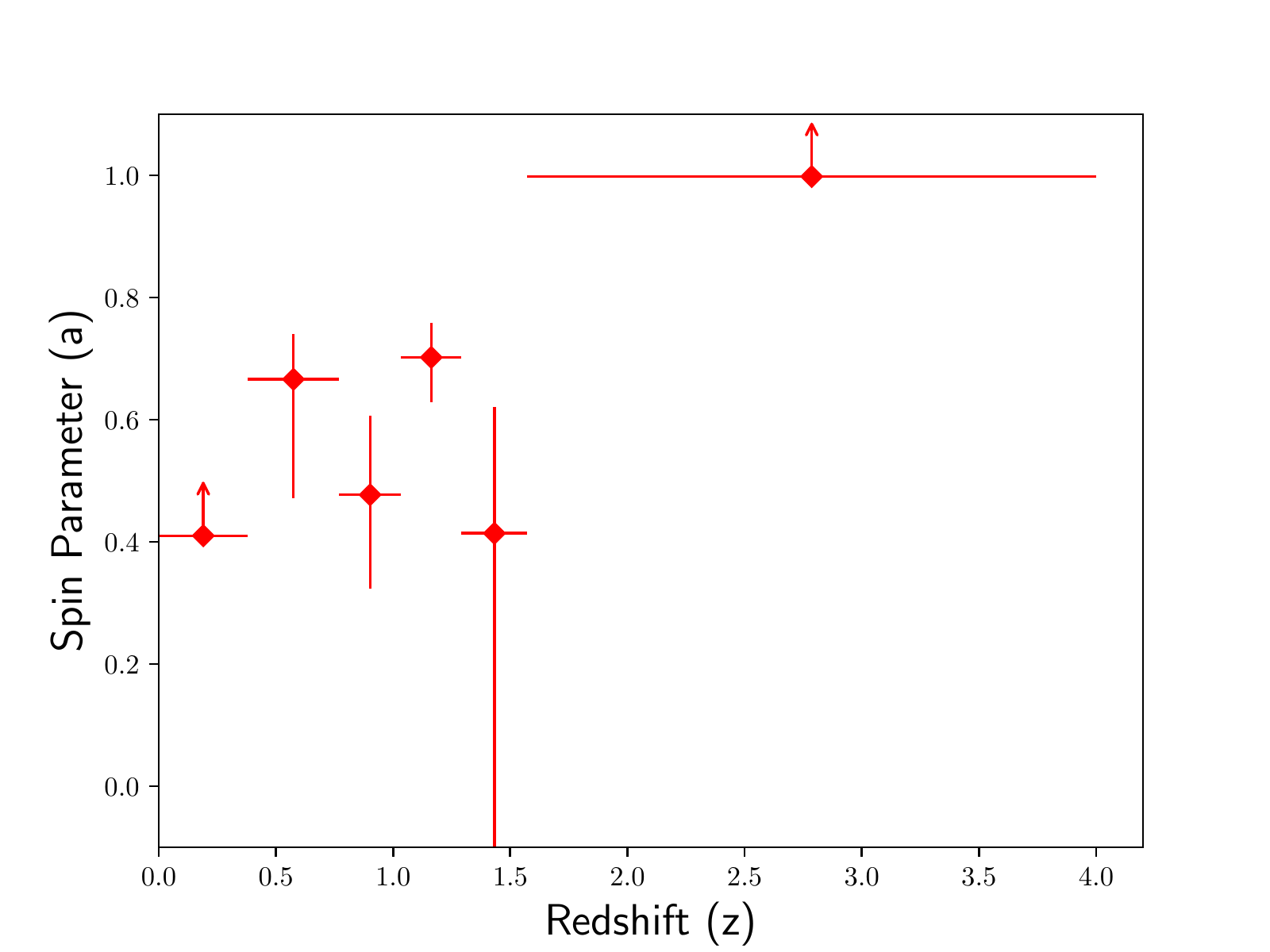}} &  \resizebox{70mm}{!}{\includegraphics{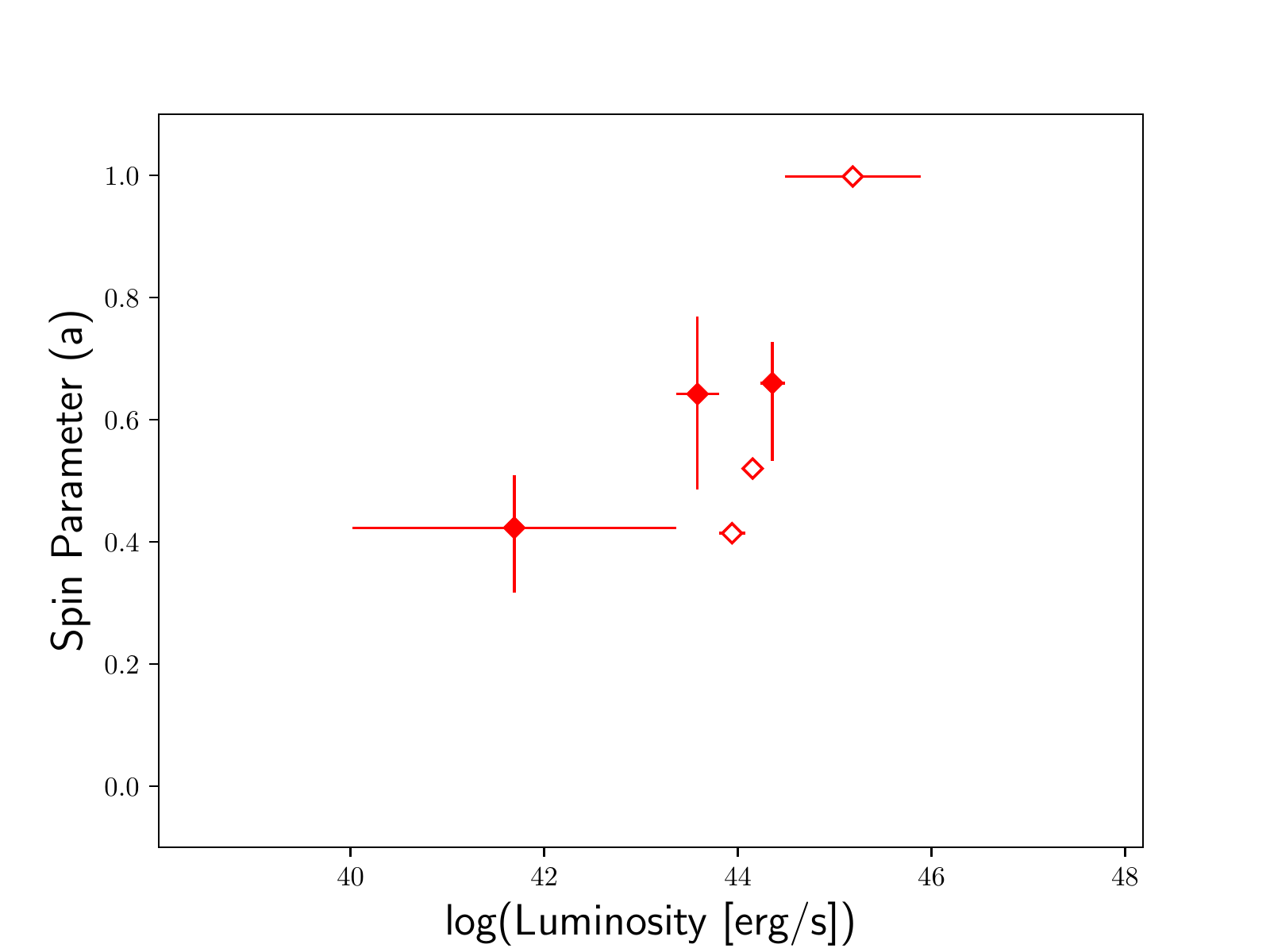}} \\
\resizebox{70mm}{!}{\includegraphics{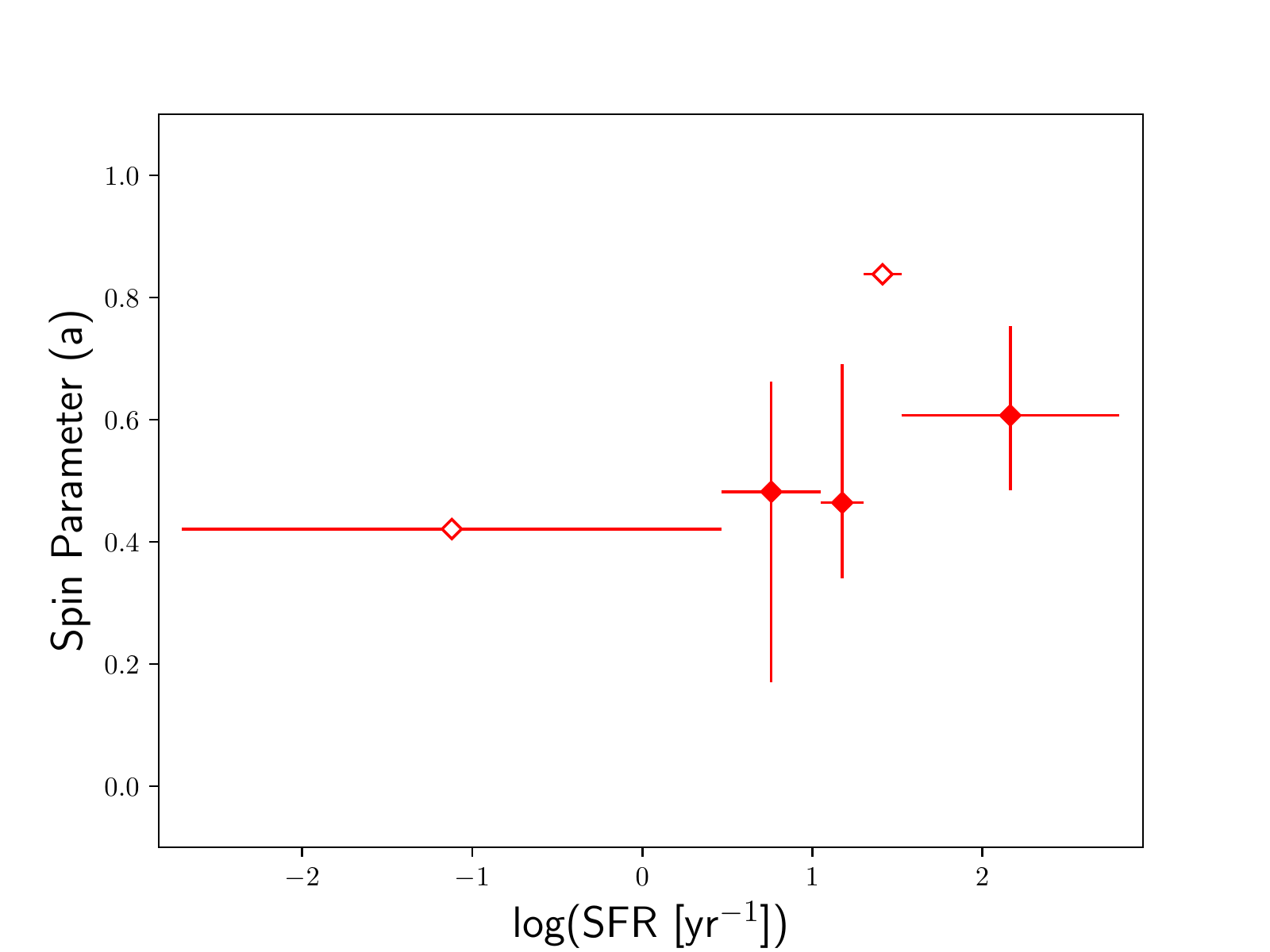}} &  \resizebox{70mm}{!}{\includegraphics{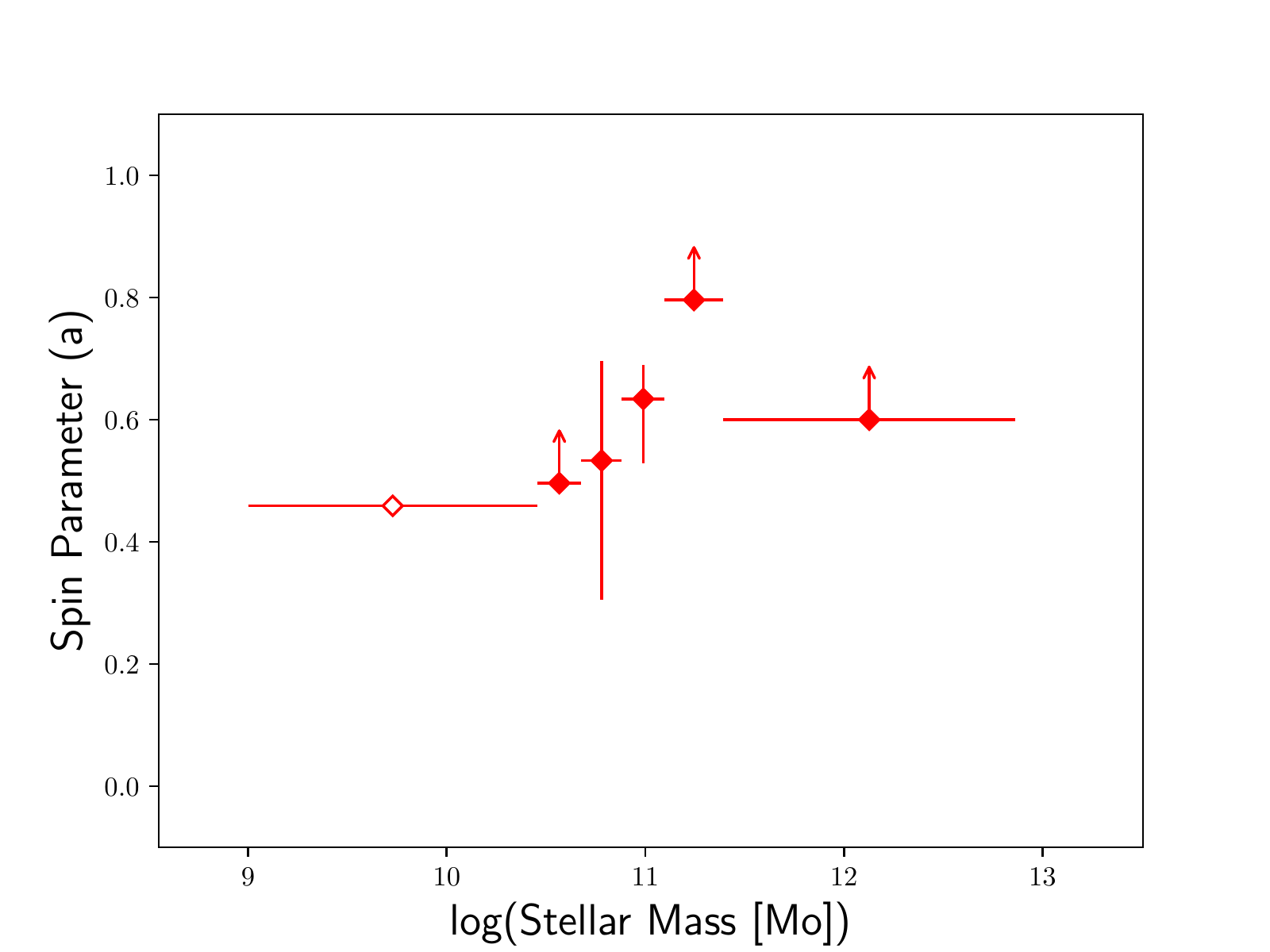}} \\
\resizebox{70mm}{!}{\includegraphics{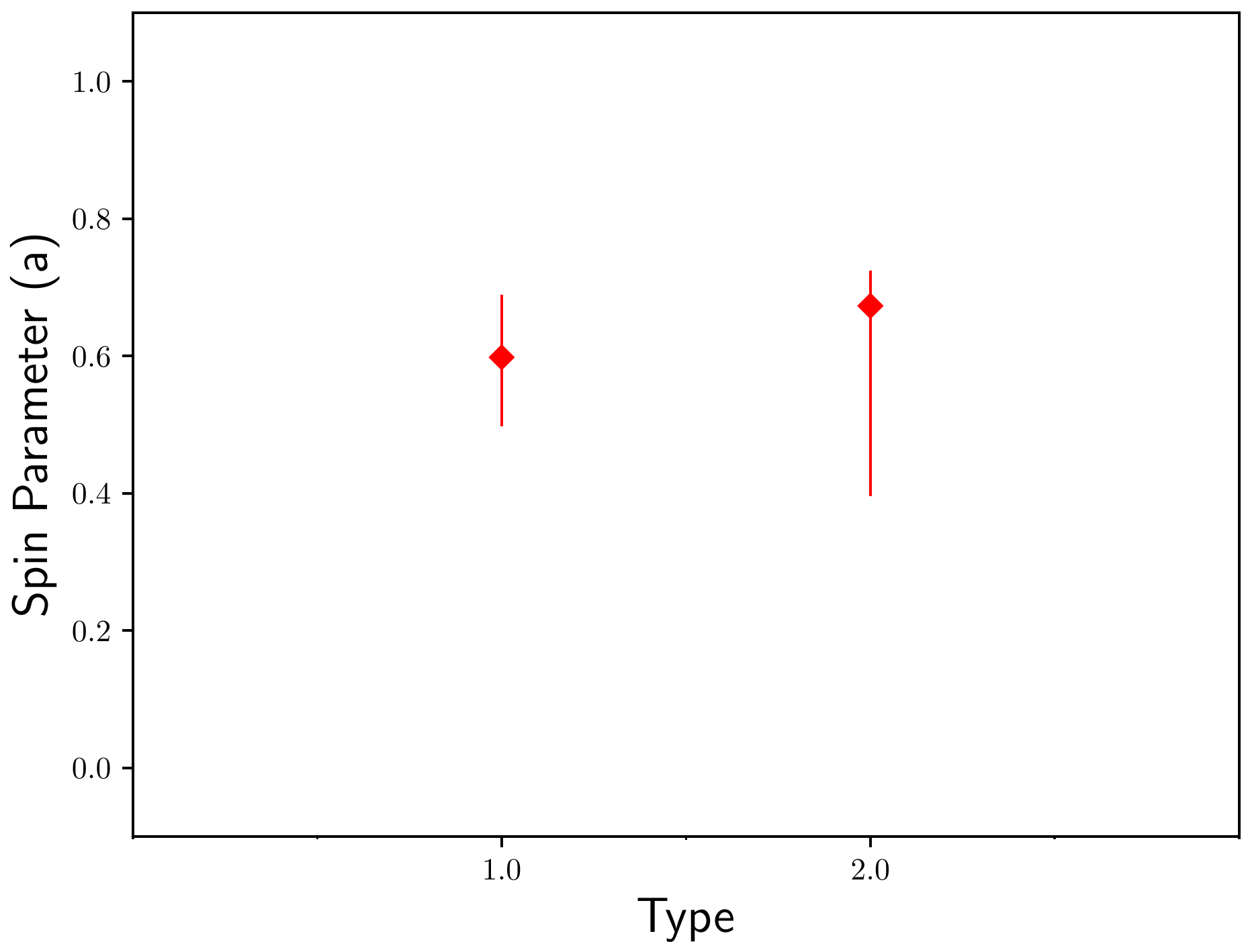}} &  \resizebox{70mm}{!}{\includegraphics{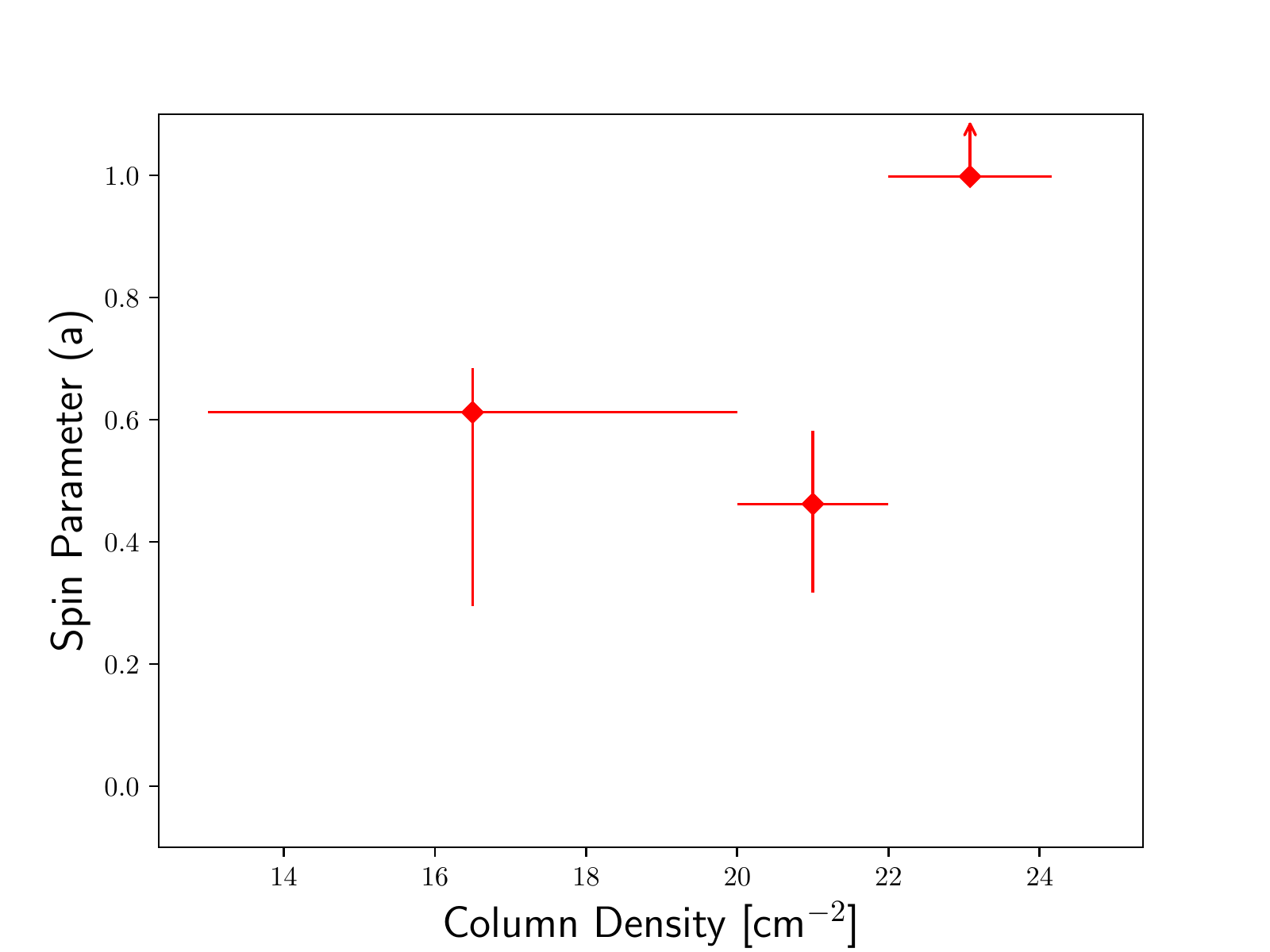}} \\
 \multicolumn{2}{c}{\resizebox{70mm}{!}{\includegraphics{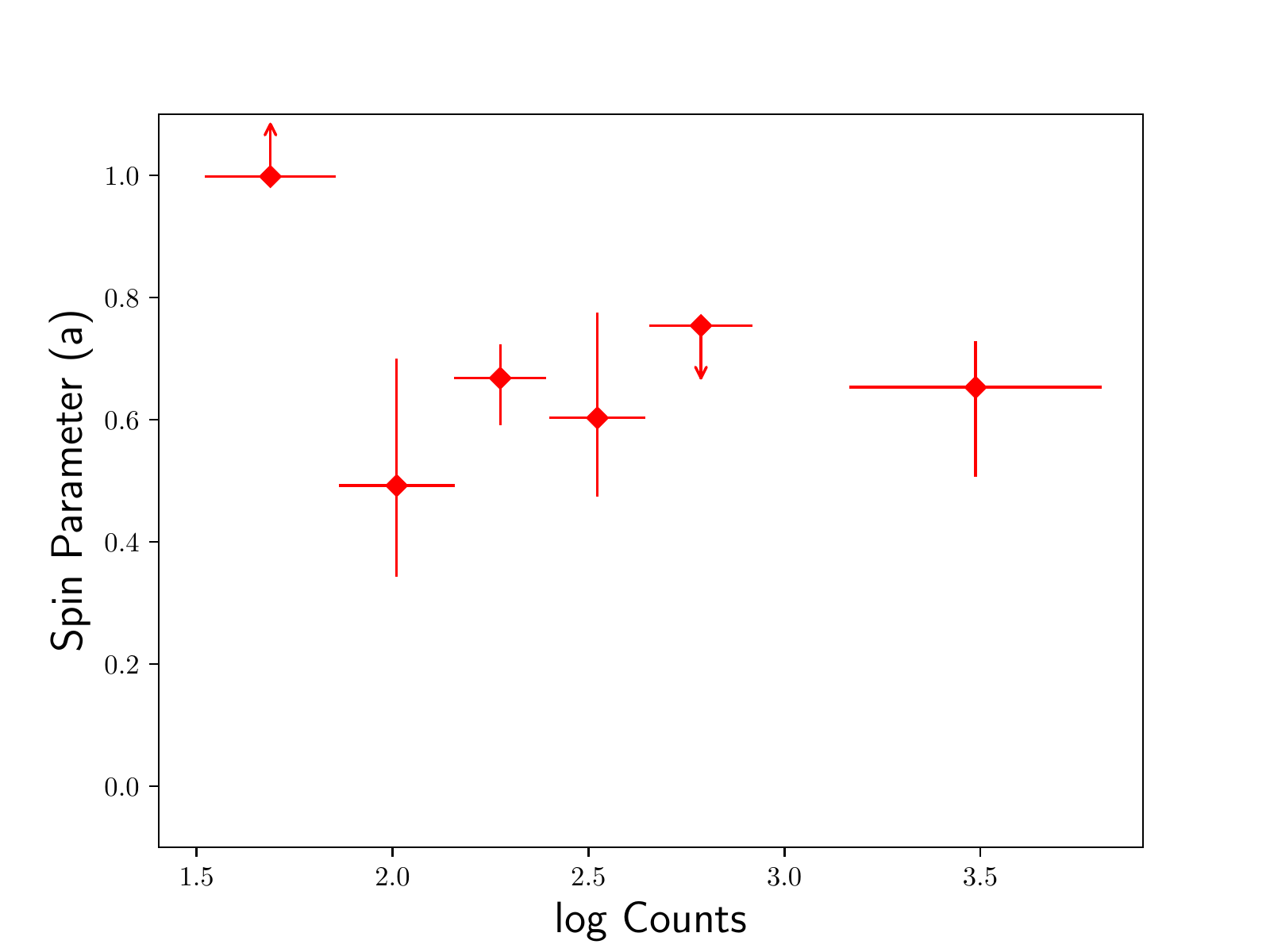}}} \\
\end{tabular}
\caption{Evolution of the best-fit \texttt{relline}+Gaussian spin parameter \textit{a} in bins of observational properties. Arrows indicate where the spin parameter represents an upper limit and is not well constrained. Empty markers indicate an unconstrained spin parameter. \label{fig:fit:rlg}}
\end{center}
\end{figure*}

\begin{table*}
\centering \footnotesize
\caption{Best-fit Gaussian widths, \texttt{relline} and \texttt{relline}+Gaussian black hole spin measurements for the full average CCLS sample. Also included are the best-fit parameters for the CCLS sample further divided by observational properties. $U$ indicates an unconstrained parameter.} \label{tab:fits}
\begin{tabular}{cccccc}
\hline\hline
 &  & \multicolumn{2}{c}{Gaussian} & \texttt{relline} & \texttt{relline}+Gaussian\\
 &  & $\sigma$ & EW & \textit{a} & \textit{a} \\ \hline
 & All Sources & $0.28\pm0.05$ & $0.13\pm0.02$ & $0.76\pm0.02$ &  $0.62~\substack{+0.07 \\ -0.17}$\\ \hline
 & Bins & $\sigma$ & EW & \textit{a}   \\ \hline
$z$ & [0.00, 0.38] & $0.18\pm0.01$ & $0.20\pm0.01$ & $<0.49$   & $>0.41$ \\
 & [0.38, 0.77] & $0.21\pm0.01$ & $0.12\pm0.01$ & $0.63\pm0.06$  & $0.67~\substack{+0.07 \\ -0.20}$ \\
 & [0.77, 1.03] & $0.28\pm0.01$ & $0.15\pm0.01$ & $0.30~\substack{+0.11 \\ -0.71}$ & $0.48~\substack{+0.13 \\ -0.15}$  \\
 & [1.03, 1.29] & $0.13\pm0.01$ & $0.12\pm0.003$ & $0.53~\substack{+0.06 \\ -0.07}$ & $0.70~\substack{+0.06 \\ -0.07}$  \\
 & [1.29, 1.57] & $0.18\pm0.02$ & $0.07\pm0.005$ & $0.53~\substack{+0.16 \\ -0.12}$ & $0.41~\substack{+0.21 \\ -1.03}$  \\
 & [1.57, 4.00] & $0.19\pm0.01$ & $0.09\pm0.004$ & $0.99~\substack{+0.001 \\ -0.03}$ & $>0.99$  \\ \hline
log $L_x$ & [40.0, 43.4] & $0.32\pm0.01$ & $0.18\pm0.01$ & $<0.33$ & $0.42~\substack{+0.09 \\ -0.11}$  \\
 & [43.4, 43.8] & $0.15\pm0.01$ & $0.11\pm0.004$ & $0.68~\substack{+0.03 \\ -0.04}$ & $0.64~\substack{+0.13 \\ -0.16}$  \\
 & [43.8, 44.1] & $0.20\pm0.01$ & $0.09\pm0.004$ & $0.67~\substack{+0.06 \\ -0.15}$ & $0.41 ~U$  \\
 & [44.1, 44.2] & $0.26\pm0.02$ & $0.15\pm0.01$ & $<0.42$ & $0.52 ~U$  \\
 & [44.2, 44.5] & $0.17\pm0.01$ & $0.11\pm0.004$ & $0.68~\substack{+0.05 \\ -0.04}$ & $0.66~\substack{+0.07 \\ -0.13}$  \\
 & [44.5, 45.9] & $0.16\pm0.01$ & $0.08\pm0.01$ & $0.75~\substack{+0.06 \\ -0.10}$ & $0.99 ~U$  \\ \hline
log $SFR$ & [-2.71, 0.47] & $0.20\pm0.02$ & $0.11\pm0.01$ & $<0.37$ & $0.42 ~U$  \\
 & [0.47, 1.05] & $0.24\pm0.01$ & $0.12\pm0.01$ & $<0.00$ & $0.48~\substack{+0.18 \\ -0.31}$  \\
 & [1.05, 1.30] & $0.18\pm0.01$ & $0.15\pm0.01$ & $0.49~\substack{+0.05 \\ -0.09}$ & $0.46~\substack{+0.23 \\ -0.12}$  \\
 & [1.30, 1.52] & $0.21\pm0.02$ & $0.08\pm0.01$ & $<0.54$ & $0.84 ~U$  \\
 & [1.52, 2.80] & $0.21\pm0.01$ & $0.11\pm0.004$ & $0.77~\substack{+0.03 \\ -0.04}$ & $0.61~\substack{+0.15 \\ -0.12}$  \\ \hline
log $M^*$ & [9.0, 10.5] & $0.15\pm0.01$ & $0.08\pm0.003$ & $0.57~\substack{+0.06 \\ -0.09}$ & $0.46 ~U$  \\
 & [10.5, 10.7] & $0.13\pm0.004$ & $0.08\pm0.003$ & $0.54~\substack{+0.12 \\ -0.09}$ & $>0.50$  \\
 & [10.7, 10.9] & $0.23\pm0.01$ & $0.13\pm0.01$ & $0.52~\substack{+0.15 \\ -0.13}$ & $0.53~\substack{+0.16 \\ -0.23}$  \\
 & [10.9, 11.1] & $0.23\pm0.02$ & $0.14\pm0.01$ & $0.60~\substack{+0.04 \\ -0.05}$ & $0.63~\substack{+0.06 \\ -0.11}$  \\
 & [11.1, 11.4] & $0.26\pm0.02$ & $0.13\pm0.01$ & $0.61~\substack{+0.06 \\ -0.15}$ & $>0.80$  \\
 & [11.4, 12.9] & $0.17\pm0.01$ & $0.12\pm0.01$ & $0.74~\substack{+0.03 \\ -0.06}$ & $>0.60$  \\ \hline
Type & 1 & $0.25\pm0.01$ & $0.11\pm0.005$ & $0.61~\substack{+0.05 \\ -0.08}$ & $0.60~\substack{+0.09 \\ -0.10}$  \\
 & 2 & $0.17\pm0.01$ & $0.10\pm0.003$ & $0.69\pm0.03$ & $0.67~\substack{+0.05 \\ -0.28}$  \\ \hline
log $N_H$ & [13.0, 20.0] & $0.27\pm0.01$ & $0.14\pm0.004$ & $0.78\pm0.03$ & $0.61~\substack{+0.07 \\ -0.32}$  \\
 & [20.0, 22.0] & $0.16\pm0.01$ & $0.09\pm0.003$ & $0.60~\substack{+0.06 \\ -0.09}$ & $0.46~\substack{+0.12 \\ -0.15}$  \\
 & [22.0, 24.2] & $0.15\pm0.01$ & $0.07\pm0.003$ & $0.67\pm0.04$ & $0.99~\substack{+0.0003 \\ -0.004}$  \\ \hline
Counts & [30, 68] & $0.20\pm0.01$ & $0.09\pm0.004$ & $0.77~\substack{+0.03 \\ -0.05}$ & $>0.99$  \\
 & [68, 137] & $0.21\pm0.01$ & $0.10\pm0.004$ & $0.66~\substack{+0.03 \\ -0.06}$ & $0.49~\substack{+0.21 \\ -0.15}$  \\
 & [137, 239] & $0.23\pm0.01$ & $0.13\pm0.01$ & $0.35~\substack{+0.08 \\ -0.10}$ & $0.67~\substack{+0.06 \\ -0.08}$  \\
 & [239, 427] & $0.19\pm0.01$ & $0.10\pm0.004$ & $0.52~\substack{+0.08 \\ -0.10}$ & $0.60~\substack{+0.17 \\ -0.13}$  \\
 & [427, 798] & $0.24\pm0.01$ & $0.11\pm0.004$ & $0.54~\substack{+0.06 \\ -0.10}$ & $<0.75$  \\
 & [798, 5355] & $0.23\pm0.01$ & $0.15\pm0.01$ & $0.72\pm0.03$ & $0.65~\substack{+0.08 \\ -0.15}$ \\ \hline
\end{tabular}
\end{table*}

\begin{table*}
\centering \footnotesize
\caption{Slope of a linear fit to the Gaussian width, \texttt{relline} and \texttt{relline}+Gaussian black hole spin parameters as a function of observational properties. The statistical significance of these slopes are given by the column $\sigma$.} \label{tab:slopes}
\begin{tabular}{crlrlrl}
\hline\hline
 & \multicolumn{2}{c}{Gaussian} & \multicolumn{2}{c}{\texttt{relline}} & \multicolumn{2}{c}{\texttt{relline}+Gaussian}\\
Parameter & \multicolumn{1}{c}{Slope} & \multicolumn{1}{c}{($\sigma$)} & \multicolumn{1}{c}{Slope} & \multicolumn{1}{c}{($\sigma$)} & \multicolumn{1}{c}{Slope} & \multicolumn{1}{c}{($\sigma$)} \\ \hline
$z$ & $-0.001\pm0.034$ & $0.03$ & $0.195\pm0.083$ & $2.3$ & $0.190\pm0.083$ & 2.3 \\
log $L_x$ & $-0.046\pm0.019$ & $2.4$ & $0.111\pm0.047$ & $2.4$ & $0.130\pm0.066$ & 2.0 \\
log $SFR$ & $0.002\pm0.013$ & $0.2$ & $0.112\pm0.115$ & $1.0$ & $0.077\pm0.066$ & 1.2 \\
log $M^*$ & $0.026\pm0.029$ & $0.9$ & $0.076\pm0.032$ & $2.4$ & $0.089\pm0.062$ & 1.4 \\
log $N_H$ & $-0.019\pm0.005$ & $3.8$ & $-0.020\pm0.018$ & $1.1$ & $0.044\pm0.070$ & 0.6 \\
Counts & $0.017\pm0.013$ & $1.3$ & $-0.003\pm0.122$ & $0.02$ & $-0.084\pm0.129$ & 0.7 \\ \hline
\end{tabular}
\end{table*}


\subsection{The Average CCLS Spin}

By stacking the CCLS sources, we are able to extract the average Fe K$\alpha$ emission line profile at 6.4 keV. We first fit this emission line with a Gaussian model and find a broadened line profile with a Gaussian width of $\sigma_g=0.28\pm0.05$ and equivalent width of $\text{EW}=0.13\pm0.02$ (Table \ref{tab:fits}). 
In a similar study with \textit{XMM}, \citet{Cor08} and \citet{Cor11} do not find compelling evidence for broad relativistic Fe K$\alpha$ emission. They do, however, place upper limits on the equivalent width of a relativistic line at $\text{EW}=0.40$ keV and $\text{EW}=0.23$ keV, respectively. 
Other stacking analyses with \textit{XMM} survey fields likewise do not find evidence for relativistically broadened Fe K$\alpha$ line emission (e.g., \citealt{Cor08,Iwa12,Fal13,Liu16}).

The width of the Fe K$\alpha$ emission line that we find suggests that the average spectra in the CCLS sample is undergoing mild relativistic broadening. This broadening, however, may be explained by a variety of processes other than a relativistically spinning black hole, including disk inclination, reflection from circumnuclear clouds, and physical processes like gas turbulence and ionization. 

To help constrain the origin of this broad Fe K$\alpha$ emission line profile, we then fit the averaged CCLS spectra with relativistic emission line models. Starting with \texttt{diskline}, a broadened emission line for a non-spinning black hole, and working up to the relativistic line model \texttt{relline} for which the black hole spin parameter, $a$, may be constrained. We find the single best-fit \texttt{relline} model yields a spin measurement of $0.76\pm0.02$, while the more complex, best-fit \texttt{relline} + Gaussian model finds $a=0.62~\substack{+0.07 \\ -0.17}$. 
These are comparable to the average prograde spin, $\text{a}\sim0.7$, found by \citet{Wal15} for 27 lensed quasars. 

The average spin we find for both relativistic models suggests that the dominant growth mechanism for this AGN population is likely merger dominated. 
Relativistic numerical simulations from \citet{Ber08} show that the average black hole spin from randomly oriented, isotropically distributed major mergers is $a\sim0.69$, which falls between our two best-fit spins. Chaotic accretion episodes cause the spin to slow, while prolonged accretion aligned with the spin direction can spin up the black hole. Thus, while major mergers likely dominate our population, there may be contributions from prograde accretion (\texttt{relline} model) and/or chaotic accretion (\texttt{relline} + Gaussian model).

\subsection{Black Hole Spin: Evolution and Environment} \label{ssec:bhs}

We take this analysis a step further by separating our average data/model ratios in bins of observed properties to investigate whether evolution and environment may be linked to the Fe K$\alpha$ emission line profile, and by extension, the spin of the SMBH (Figures \ref{fig:fit:gauss}, \ref{fig:fit:relline}; Tables \ref{tab:fits}, \ref{tab:slopes}). The relativistic line models used in the following analysis are based on the best-fit parameters for the single \texttt{relline} model, and the more complex \texttt{relline} + Gaussian models. In order to constrain the spin parameter in the complex model for these smaller bins, however, all parameters, excluding the normalization and spin, had to be frozen to their best-fit values. We fit a line to our observed parameter bins using a non-linear least-squares optimization with scipy \texttt{curve\_fit} to provide a rough estimate of the strength of any possible parameter evolution.

\subsubsection{Evolution to $z\sim5.3$}

Binning by redshift, we find no evidence for evolution of the Gaussian width (Figure \ref{fig:fit:gauss}; top, left), best fit with a slope of $m_z=-0.001\pm0.034$ (slope significance; $\sigma=0.03$). The black hole spin parameters extracted have a stronger slope than the Gaussian widths for both the single relativistic line and the more complex \texttt{relline} + Gaussian model (Table \ref{tab:slopes}), and these slopes are consistent (\texttt{relline}: $m_z=0.195\pm0.083$, $\sigma=2.3$; \texttt{relline}+Gaussian: $m_z=0.190\pm0.083$, $\sigma=2.3$).
However, this fit is heavily influenced by the highest redshift bin where sources are fainter and less abundant. Excluding this high-z bin, the spin parameter does not appear to evolve with redshift.

\subsubsection{Luminosity}

Interestingly, we find anti-correlated dependencies of the Gaussian width and black hole spin parameter in bins of $2-10$ keV luminosity (Figure \ref{fig:fit:gauss}, \ref{fig:fit:relline}; top, right). The Gaussian widths evolve according to a best-fit slope of $m_{Lx}=-0.046\pm0.019$ ($\sigma=2.4$). The black hole spin best-fit slopes in comparison are $m_{Lx}=0.111\pm0.047$ ($\sigma=2.4$), and $m_{Lx}=0.130\pm0.066$ ($\sigma=2.0$) for the single \texttt{relline} and \texttt{relline} + Gaussian models, respectively. These fits, however, are predominately influenced by the lowest luminosity bin. Furthermore, 50\% of the extracted spins are unconstrained for the more complex relativistic model. Perhaps unsurprisingly, the Gaussian widths (and equivalent widths; Figure \ref{fig:ew}) exhibit an inverse relationship that is consistent with the X-ray Baldwin effect (the inverse correlation between the equivalent width and $L_X$; \citealt{Iwa93}). The black hole spin parameters, however, increase with increasing luminosity. 
This may reflect the interesting impact that spin has on the AGN Eddington limit. For faster spinning black holes, the ISCO moves inward toward the event horizon, enabling stable orbits deeper within the gravitational potential well, and thus enabling more energy to be extracted from inflating material, increasing the AGN's capacity to be more luminous.

\subsubsection{Stellar Mass and SFR}

From observational evidence, it has become increasingly clear that there exists a correlation between the growth of the SMBH and its host galaxy (\citealt{Fer00}), however, there is disagreement about which galaxy property ($SFR$ or $M^*$) is most fundamentally linked with AGN activity (e.g., \citealt{Che13,Yan18,Yan19,Suh19,Ste20}). 
By binning the CCLS sample by $SFR$ and $M^*$, we may test the connection between these galaxy properties and the shape of the Fe K$\alpha$ line profile, and by extension, the AGN activity (Figure \ref{fig:fit:gauss}, \ref{fig:fit:relline}; second row, left and right, respectively). 

For our Gaussian fits, we find no evidence linking an evolution of $SFR$ with broadened Fe K$\alpha$ ($m_{SFR}=0.002\pm0.013$; $\sigma=0.2$), while there does exist slightly more compelling evidence for a connection between $M^*$ and broadened Fe K$\alpha$ ($m_{M^*}=0.026\pm0.029$; $\sigma=0.9$), albeit not statistically significant. This does appear to be strongly influenced by the most massive bin, where fewer high mass sources may unduly skew the result. 

The \texttt{relline} best-fit spin parameters exhibit stronger positive slopes for both $SFR$ ($m_{SFR}=0.112\pm0.115$; $\sigma=1.0$) and $M^*$ ($m_{SFR}=0.076\pm0.032$; $\sigma=2.4$) compared to the Gaussian evolution, however, the spin parameters in the $SFR$ bins are not well constrained: $60\%$ are upper limits. 
The \texttt{relline} + Gaussian best-fit spin parameter slopes for $SFR$ and $M^*$ are more consistent than the single \texttt{relline} model with $m_{SFR}=0.077\pm0.066$ ($\sigma=1.2$), and $m_{SFR}=0.089\pm0.062$ ($\sigma=1.4$), respectively.
The relationship between $M^*$ and black hole spin is not quite significant for the single \texttt{relline} model, but could point to stellar mass being the predominate driver of the observed connection between the growth of the black hole and its host galaxy.

Neither of the relativistic models exhibit an anti-correlation between the spin parameter and stellar mass, that (assuming stellar mass is a good proxy for black hole mass) is exhibited by the few dozen currently known black hole spins (See \citealt{Rey19} Figure 3). While it is expected that the AGN population with known black hole spins suffers from selection biases and low number statistics that may unduly influence this observed relationship, \citet{Pac20} theoretically predict a similar trend wherein the spin parameter decreases with increasing black hole mass. Thus, our conclusions must be treated cautiously.

\subsubsection{Obscuration and AGN Type}

We compare the evolution of the Gaussian width and black hole spin parameter with the optically defined AGN type (Figure \ref{fig:fit:gauss}, \ref{fig:fit:relline}; third row, left), and obscuring column density (Figure \ref{fig:fit:gauss}, \ref{fig:fit:relline}; third row, right), as they are both commonly used to select AGN as obscured or unobscured. 
There is some evidence that these two methods may be selecting slightly different AGN populations (e.g., \citealt{Mer14}), however, \citet{Civ16} compared the selection of AGN by obscuration and Type in \textit{Chandra}-COSMOS, and found a good match between the two methods (see also \citealt{Mar16X} for CCLS). 

The Gaussian widths of the ``unobscured'' sources broaden as they become more ``obscured'' for both AGN type and obscuring column density, and the width values are consistent across both properties. This inverse relationship for column density is best-fit with a slope of $m_{NH}=-0.019\pm0.005$ ($\sigma=3.8$). 

There appears to be a comparable trend for column density with black hole spin parameter, albeit less significant ($m_{NH}=-0.020\pm0.018$; $\sigma=1.1$) for the single \texttt{relline} model. The highest column density bin in the \texttt{relline} + Gaussian model is not well constrained and significantly impacts the slope calculated for these best-fit parameters. Without this bin, the relationship between spin and column density would be more consistent with that of the single relativistic line model.
AGN type for both relativistic models, however, does not exhibit this inverse relationship with black hole spin, although the slope of the \texttt{relline} + Gaussian model is less significant than the single model.

It is possible that the inverse relationship with column density exhibited by both the Gaussian and relativistic models is influenced by a potentially large number of hidden, obscured sources where the \citet{Mar16X} spectral fits failed to properly measure $\Gamma$ and $N_H$. Observational evidence for this comes from measurements of the line EW, in which large Fe K$\alpha$ EW are an indicator of large, even CT, obscuration. The EW of our $\log N_H < 20$ cm$^{-2}$ bin is significantly higher than the column density bins with higher obscuration (Table \ref{tab:fits}).

\subsubsection{Total $0.5-7.0$ keV Counts}

We break down our data/model ratios as a function of source $0.5-7$ keV total counts to test that we are not biasing our results with the highest count sources (Figure \ref{fig:fit:gauss}, \ref{fig:fit:relline}; bottom). 

We first fit the Gaussian width as a function of counts and find a slope of $m_{cts}=0.017\pm0.013$ ($\sigma=1.3$). The flatness of this slope suggests that the S/N from a single source does not unduly influence the results of our stacking method. Similarly, we do not find evolution of the black hole spin parameter with counts for the single \texttt{relline} model ($m_{cts}=-0.003\pm0.122$; $\sigma=0.02$). There is a stronger relationship for the slope of the best-fit \texttt{relline} + Gaussian model spin parameters ($m_{cts}=-0.084\pm0.129$; $\sigma=0.02$), however, this is significantly influenced by a single bin at low counts where we expect the stacked data/model ratios to be considerably noisier.

\section{Conclusions}\label{sec:sum}

The CCLS is a rich multiwavelength environment for probing the connection between the AGN and its host galaxy. In this work, we investigated the connection between observed properties and the profiles of the Fe K$\alpha$ line emission, with implications for the average spin of SMBH across cosmic time and AGN population. 

We select all CCLS sources with $>30$ net counts in $0.5-7.0$ keV. For each of these sources we regroup the spectra from 2 to 8 keV in 100 eV bins (rest-frame). Using the best-fit spectral models from \citet{Mar16X}, we fit the continuum and calculate the data/model ratios to isolate the Fe K$\alpha$ emission line at 6.4 keV. These ratios are then averaged over the entire sample. We further divide the sample by observational properties and average these bins to test the influence of these properties on the Fe K$\alpha$ line profile.

\begin{enumerate}

\item We find the average CCLS spectrum has broad Fe K$\alpha$ line emission with Gaussian width $\sigma_g=0.28\pm0.05$ and equivalent width $EW=0.13\pm0.02$. The broad Fe K$\alpha$ line emission may also be fit with a broadened relativistic emission line with black hole spin parameter $a=0.76\pm0.02$, or better fit with a more complex \texttt{relline} + Gaussian model with $a=0.62~\substack{+0.07 \\ -0.17}$.
These black hole spin estimates are consistent with the results from \citet{Wal15} for lensed quasars. The dominant growth mechanism for the CCLS population is likely major mergers, with perhaps a smaller contribution from chaotic accretion and/or prograde accretion events (\citealt{Ber08}).

\item When dividing our sample by observational properties, we find little evolution of the Gaussian line width or black hole spin with redshift (although this conclusion excludes the highest redshift bin). Furthermore, we find no correlation between these measurements with counts in $0.5-7.0$ keV. This demonstrates that the method we are using is not biased by the highest count sources.

\item We find an interesting trend with luminosity such that the black hole spin parameter increases with increasing luminosity, while the Gaussian width (and equivalent width) are inversely correlated with increasing luminosity, consistent with the X-ray Baldwin effect. The positive trend observed with spin parameter and luminosity may be an effect from the improved accretion stability afforded by a spinning black hole, increasing the capacity of sources to become more luminous.

\item We find a possible trend in the Fe K$\alpha$ line width with stellar mass, but less so for $SFR$. This may suggest that stellar mass is the fundamental property driving the observed link between black hole and galaxy growth. Black hole spin, however, exhibits a stronger trend in $SFR$ compared to $M^*$, but the spin parameters are not well constrained in this case. Assuming stellar mass is a good proxy for black hole mass, the relationship we find is opposite to what is observed (and theorized; \citealt{Pac20}) for the few dozen nearby sources for which black hole spins have been measured (\citealt{Rey19}).

\item We find a connection between obscuration and the Fe K$\alpha$ line width and black hole spin parameter, which is consistent for both optically classified AGN types and obscuration based on X-ray spectral fits of obscuring column density. Unobscured sources (optical Type 1; $\log N_H<20$ cm$^{-2}$) have broader Fe K$\alpha$ emission line widths compared to the more obscured sources. Likewise, unobscured sources have higher black hole spins than obscured sources, although this trend is much less significant. This result may be influenced by a potentially significant, heavily obscured AGN population ``hiding'' amongst the unobscured sources.

\end{enumerate}

This work demonstrates the advantages of using \textit{Chandra} for AGN population studies. The availability of rich, deep multiwavlength fields and high spectral resolution allow for production of statistically significant samples that can be used to investigate otherwise challenging observed properties (e.g., black hole spin) and to uniquely probe the black hole-galaxy connection. 
As we look toward the future and plan the next generation of great observatories (e.g., \textit{eROSITA}, \textit{Athena}, \textit{STROBE-X}, and \textit{Lynx}), focusing on wide area, high spatial resolution instruments will be crucial to our understanding of the AGN population.

\acknowledgments

This work makes use of CIAO and Sherpa, developed by the \textit{Chandra} X-ray Center; SAOImage ds9; XSPEC, developed by HEASARC at NASA-GSFC; and the Astrophysics Data System (ADS). This work was supported by the \textit{Chandra} Archive program, grant no. AR8-19013X (PI: L. Brenneman).

\begin{appendix}

\section{Fe K$\alpha$ Line Fitting Continued} \label{app:fit}

\subsection{Gaussian Line}\label{app:gauss}

\begin{figure}
\begin{center}
\resizebox{78mm}{!}{\includegraphics{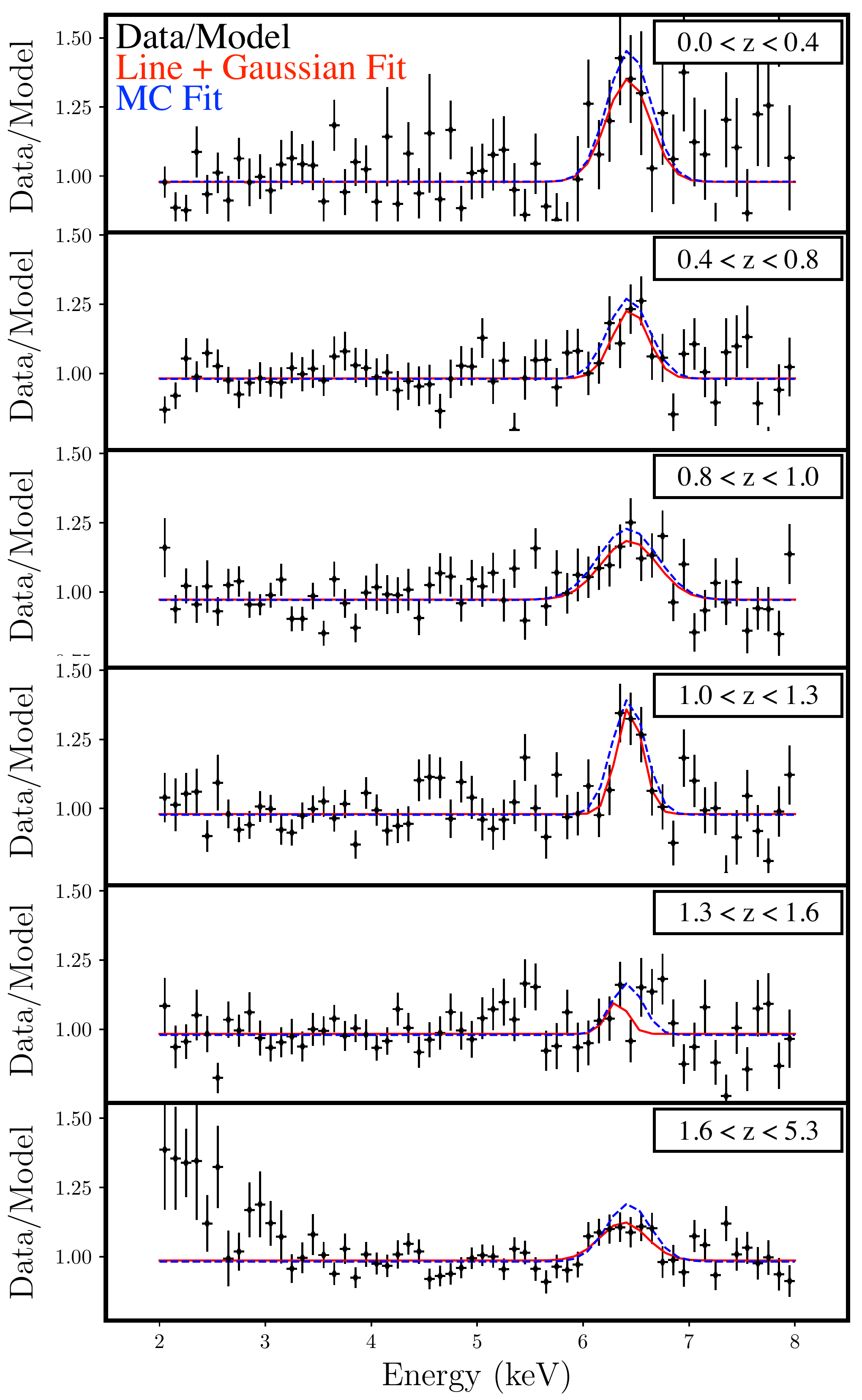}} \\
\caption{Data/Model ratio broken into bins of redshift. \label{fig:fit:z}}
\end{center}
\end{figure}

\begin{figure}
\begin{center}
\resizebox{78mm}{!}{\includegraphics{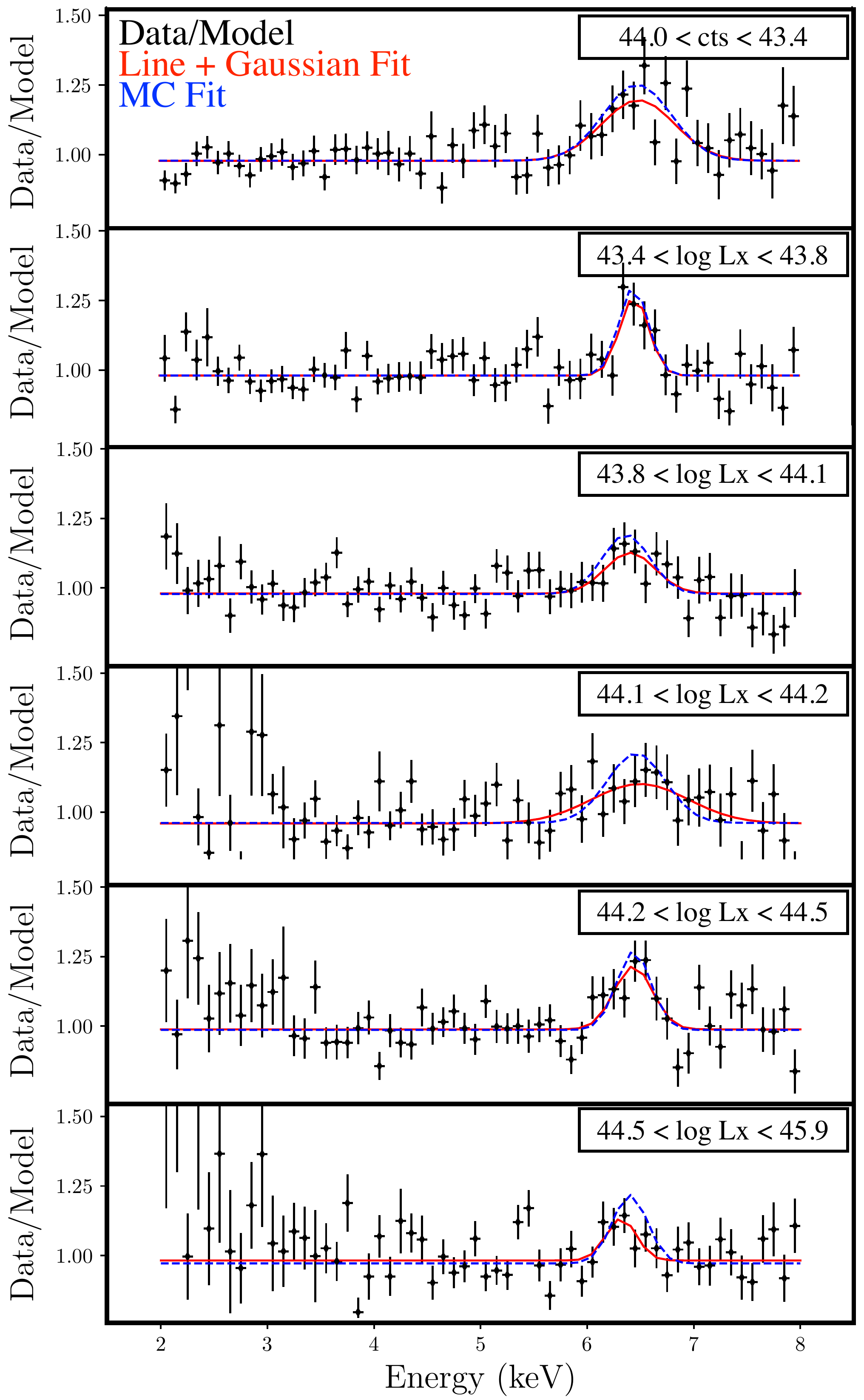}} \\
\caption{Data/Model ratio broken into bins of $2-10$ keV X-ray luminosity. \label{fig:fit:lx}}
\end{center}
\end{figure}

\begin{figure}
\begin{center}
\resizebox{78mm}{!}{\includegraphics{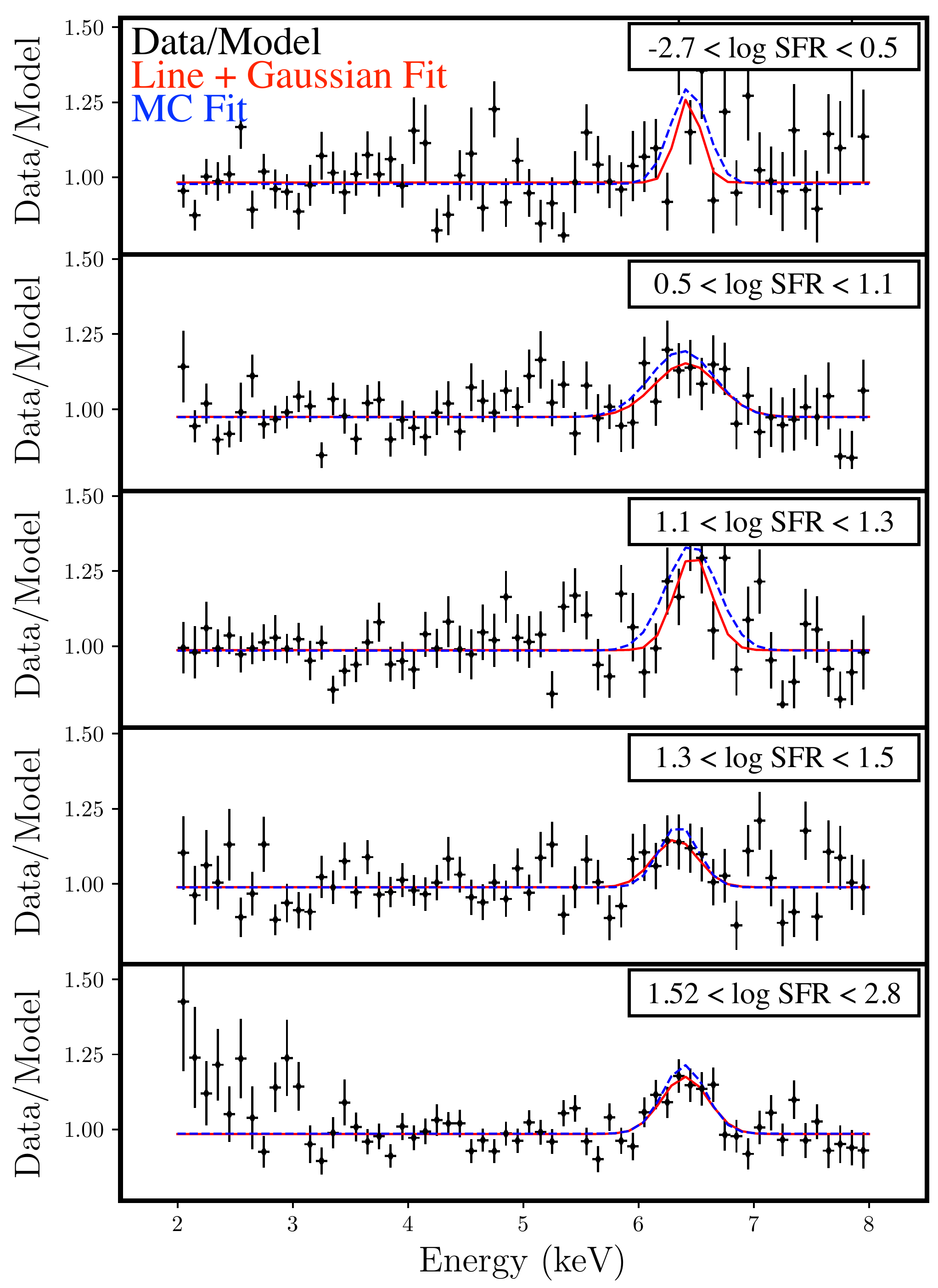}} \\
\caption{Data/Model ratio broken into bins of star formation rate. \label{fig:fit:sfr}}
\end{center}
\end{figure}

\begin{figure}
\begin{center}
\resizebox{78mm}{!}{\includegraphics{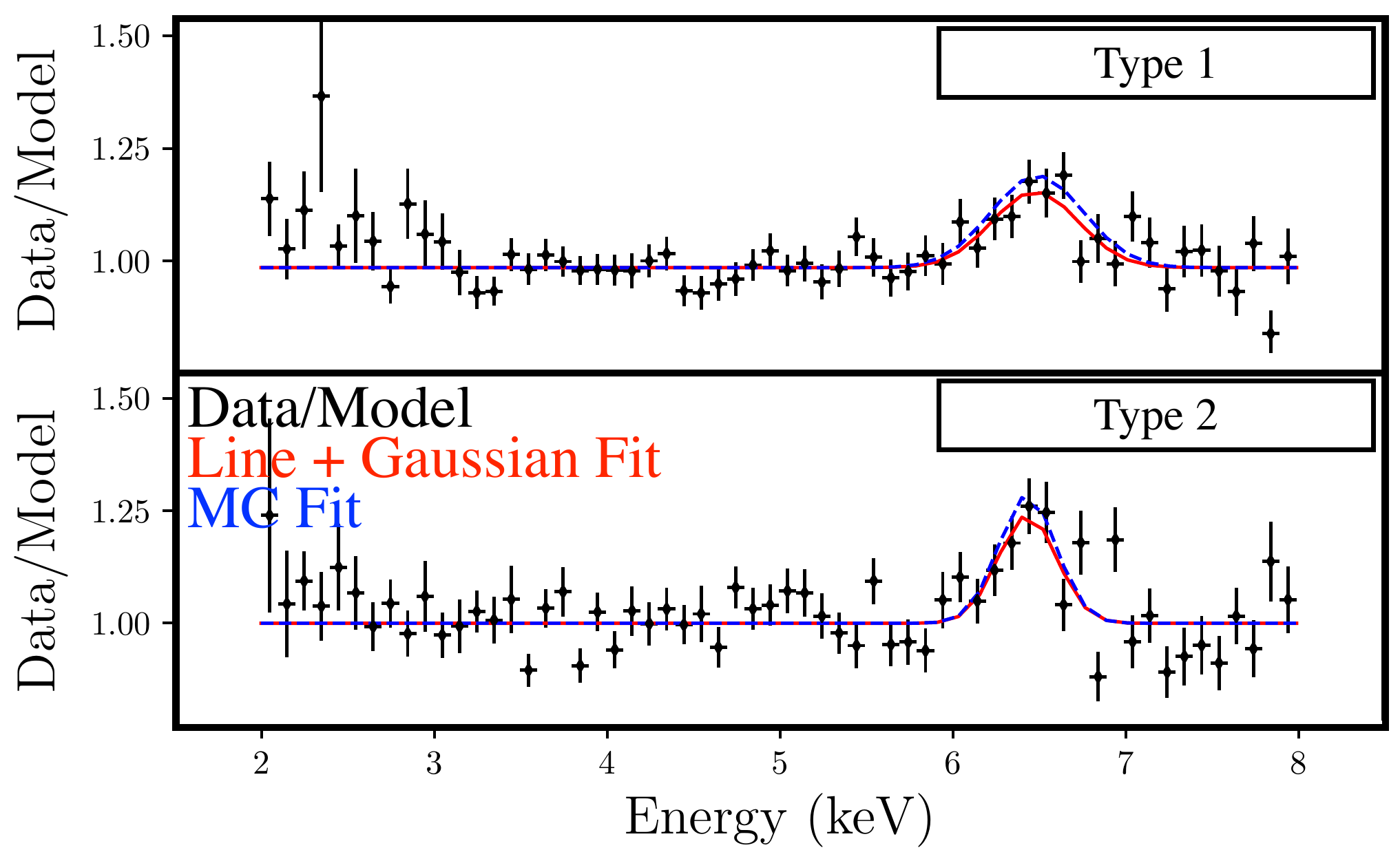}} \\
\caption{Data/Model ratio broken into bins of AGN Type. \label{fig:fit:t}}
\end{center}
\end{figure}

\begin{figure}
\begin{center}
\resizebox{78mm}{!}{\includegraphics{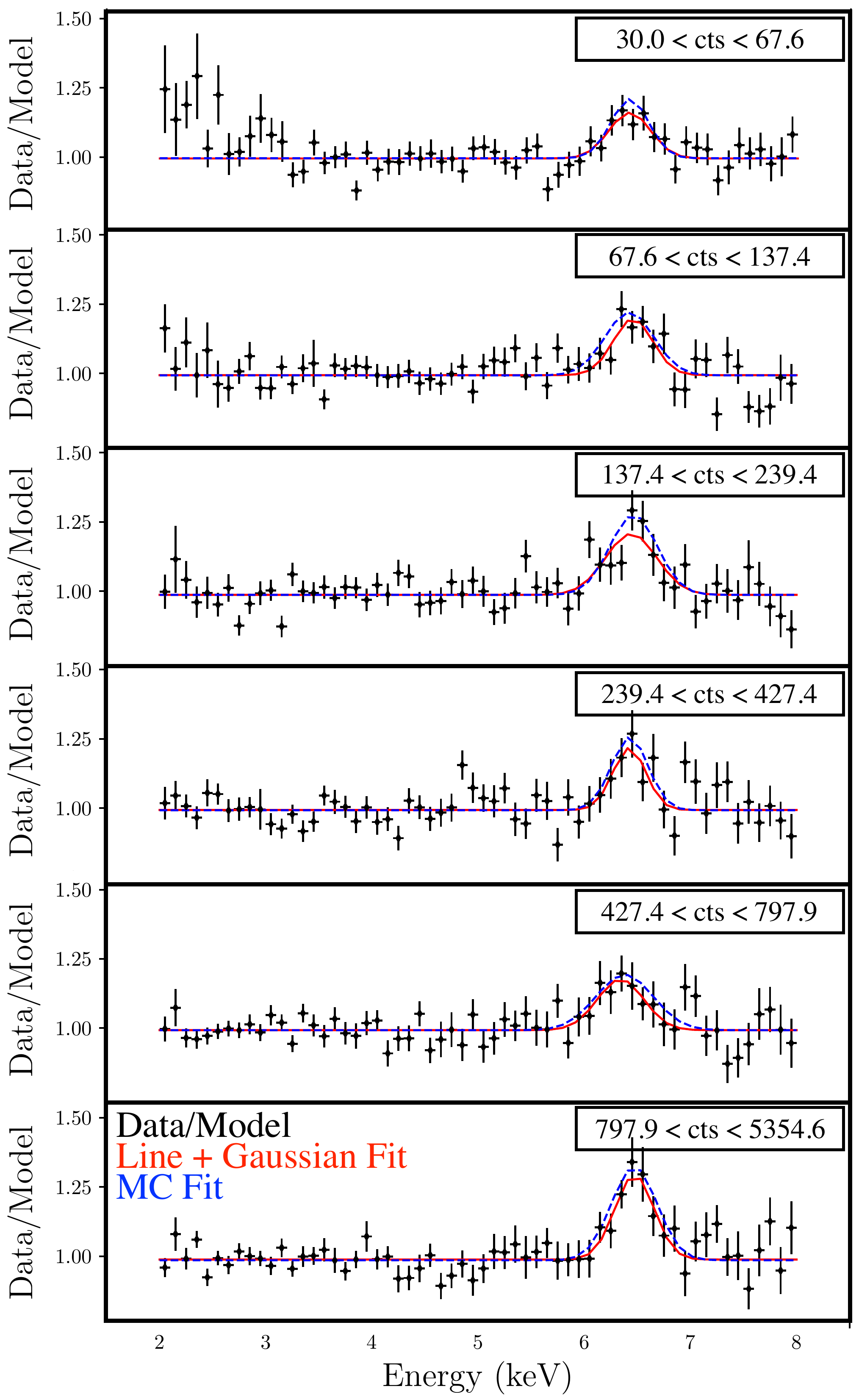}} \\
\caption{Data/Model ratio broken into bins of counts. \label{fig:fit:ct}}
\end{center}
\end{figure}

We have appended the data/model ratio Gaussian fits for each stacked bin for the host galaxy and black hole properties in our analysis: redshift (Figure \ref{fig:fit:z}), X-ray luminosity (Figure \ref{fig:fit:lx}), star formation rate (Figure \ref{fig:fit:sfr}), AGN type (Figure \ref{fig:fit:t}), and Counts (Figure \ref{fig:fit:ct}). For each data/model ratio, we adopt the Monte Carlo (N=100) best-fit parameters. These parameters are listed in Table \ref{tab:fits}.

\subsection{Relativistic Broadened Line}\label{app:rel}

\begin{figure*}
\begin{center}
\resizebox{150mm}{!}{\includegraphics{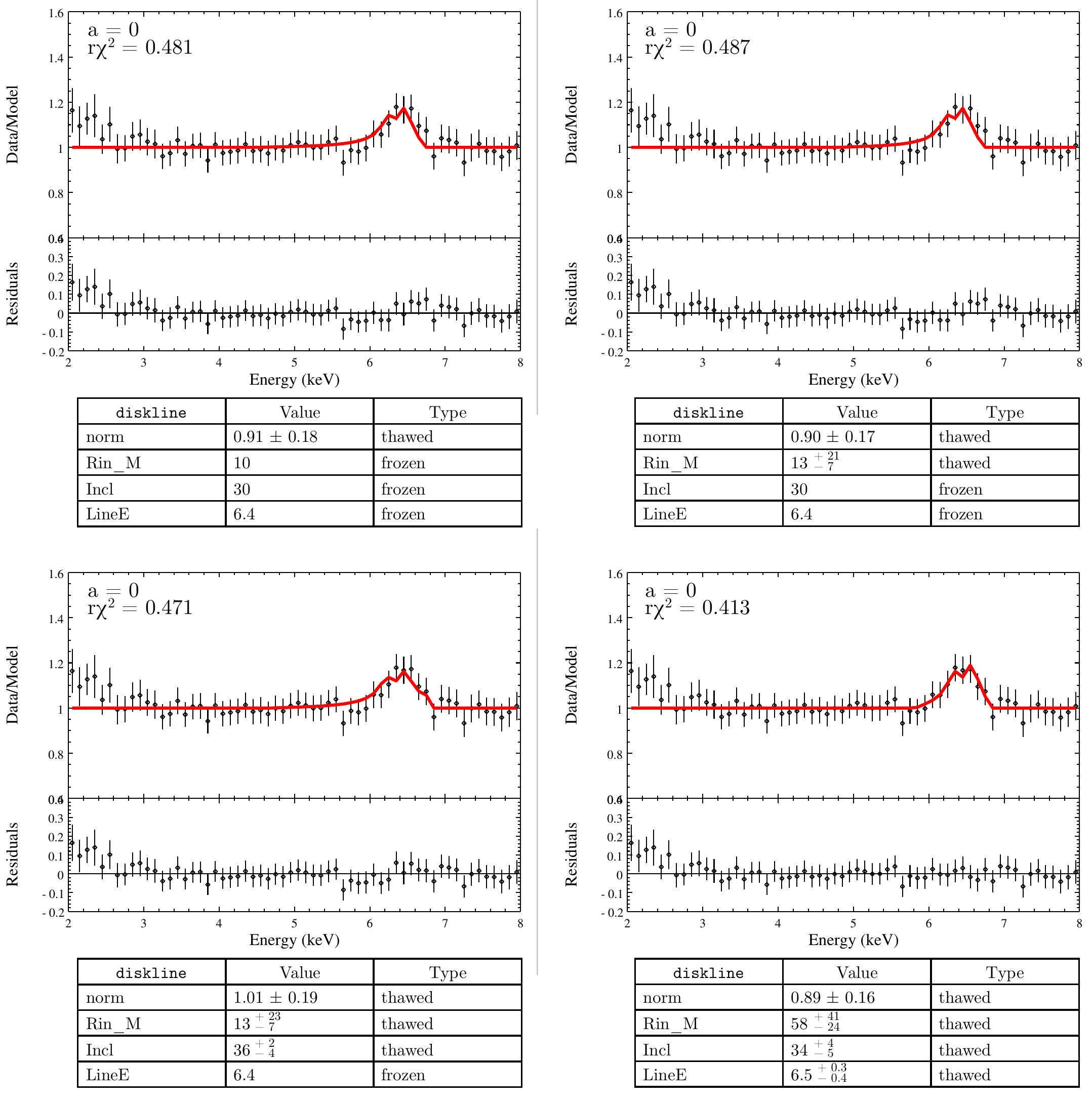}} \\
\caption{\texttt{diskline} model fits ($a=0$) to the average data/model ratio for all CCLS sources. The line energy is fixed at 6.4 keV, while all other fixed parameters are frozen at default values, unless otherwise indicated. We sequentially thaw parameters to probe this \texttt{diskline} model: (top, left) normalization thawed; (top, right) normalization and inner radius (Rin\_M) thawed; (bottom, left) normalization, inner radius, and inclination thawed; (bottom, right) normalization, inner radius, inclination, and line energy thawed. \label{fig:fit:disk}}
\end{center}
\end{figure*}

\begin{figure*}
\begin{center}
\resizebox{150mm}{!}{\includegraphics{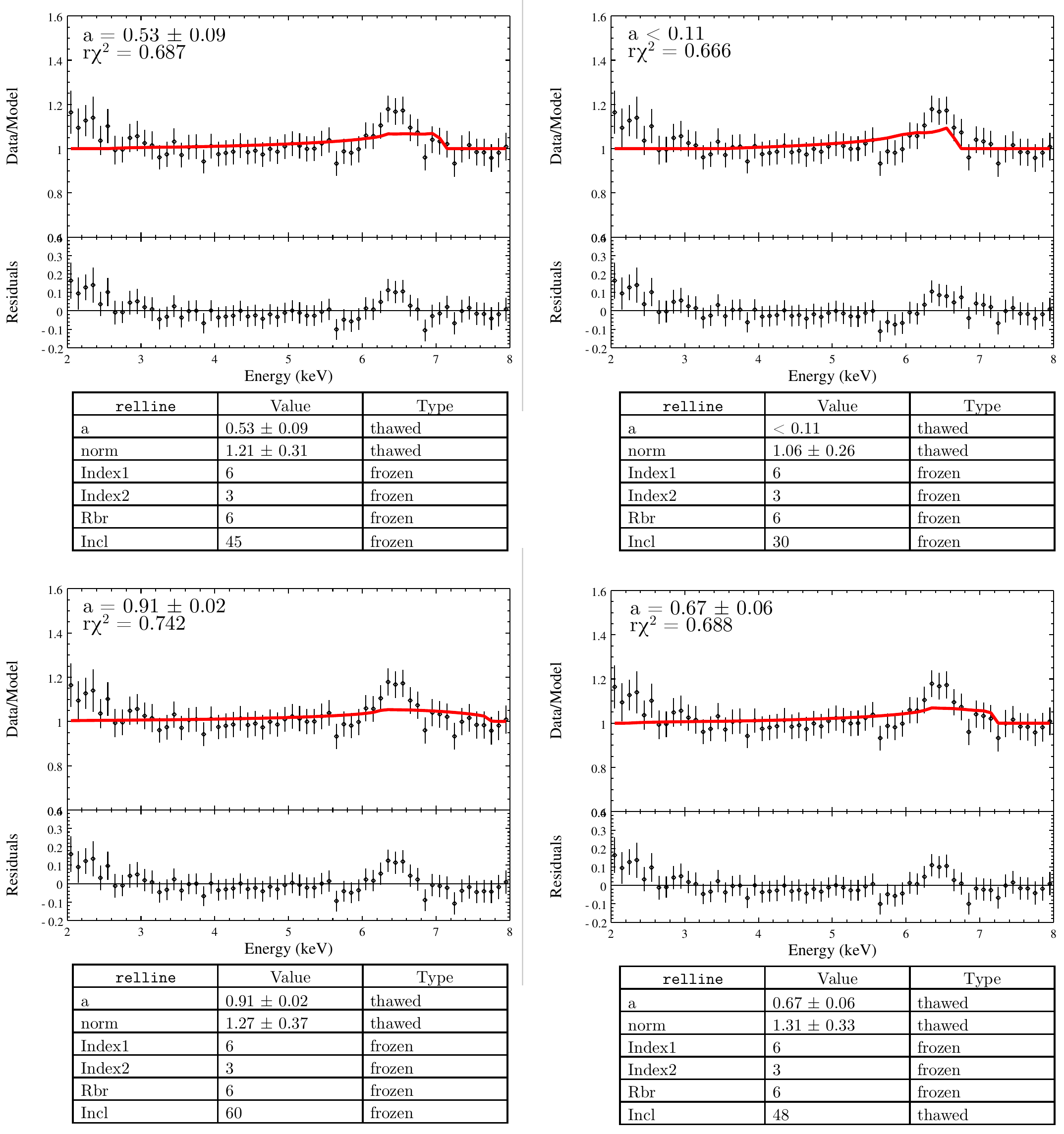}} \\
\caption{\texttt{relline} model fits to the average data/model ratio for all CCLS sources. The line energy is fixed at 6.4 keV while all other fixed parameters are frozen at default values, unless otherwise indicated. The spin parameter (a) and the normalization are left free to vary while the inclination is fixed to: (top, left) 45 degrees; (top, right) 30 degrees; (bottom, left) 60 degrees. We also allow the inclination to vary (bottom, right).\label{fig:fit:rlincl}}
\end{center}
\end{figure*}

\begin{figure*}
\begin{center}
\resizebox{150mm}{!}{\includegraphics{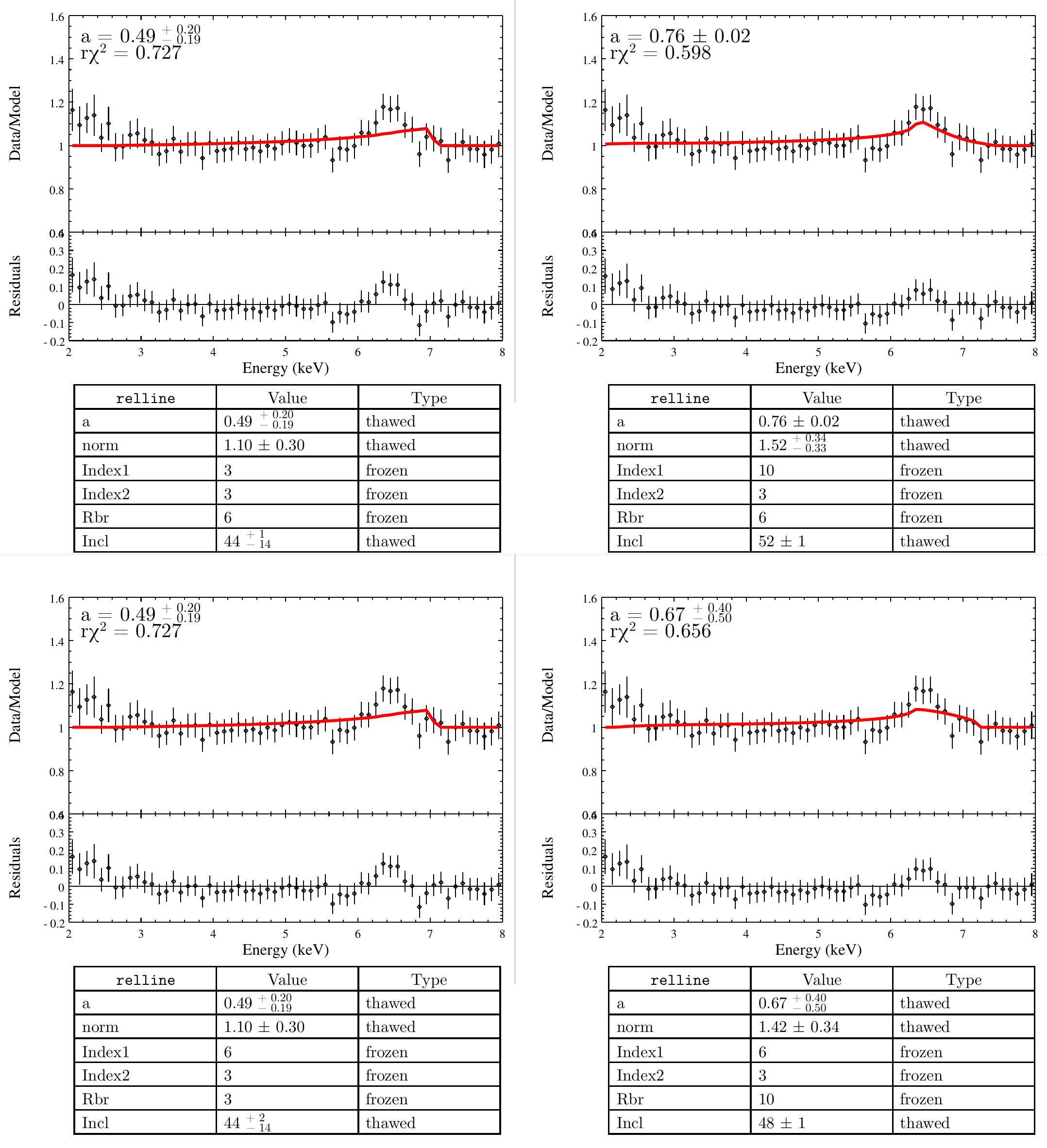}} \\
\caption{\texttt{relline} model fits to the average data/model ratio for all CCLS sources. The line energy is fixed at 6.4 keV while all other fixed parameters are frozen at default values, unless otherwise indicated. The \texttt{relline} spin parameter (a), the normalization, and the inclination are left free to vary. We explore the parameter space of the emissivity index for r $<$ Rbr (Index1) and the radius of emissivity change (Rbr) both above and below their typical values of 6: (top, left) Index1 $=3$; (top, right) Index1 $=10$; (bottom, left) Rbr $=3$; (bottom, right) Rbr = 10. \label{fig:fit:rlparam}}
\end{center}
\end{figure*}

\begin{figure*}
\begin{center}
\resizebox{150mm}{!}{\includegraphics{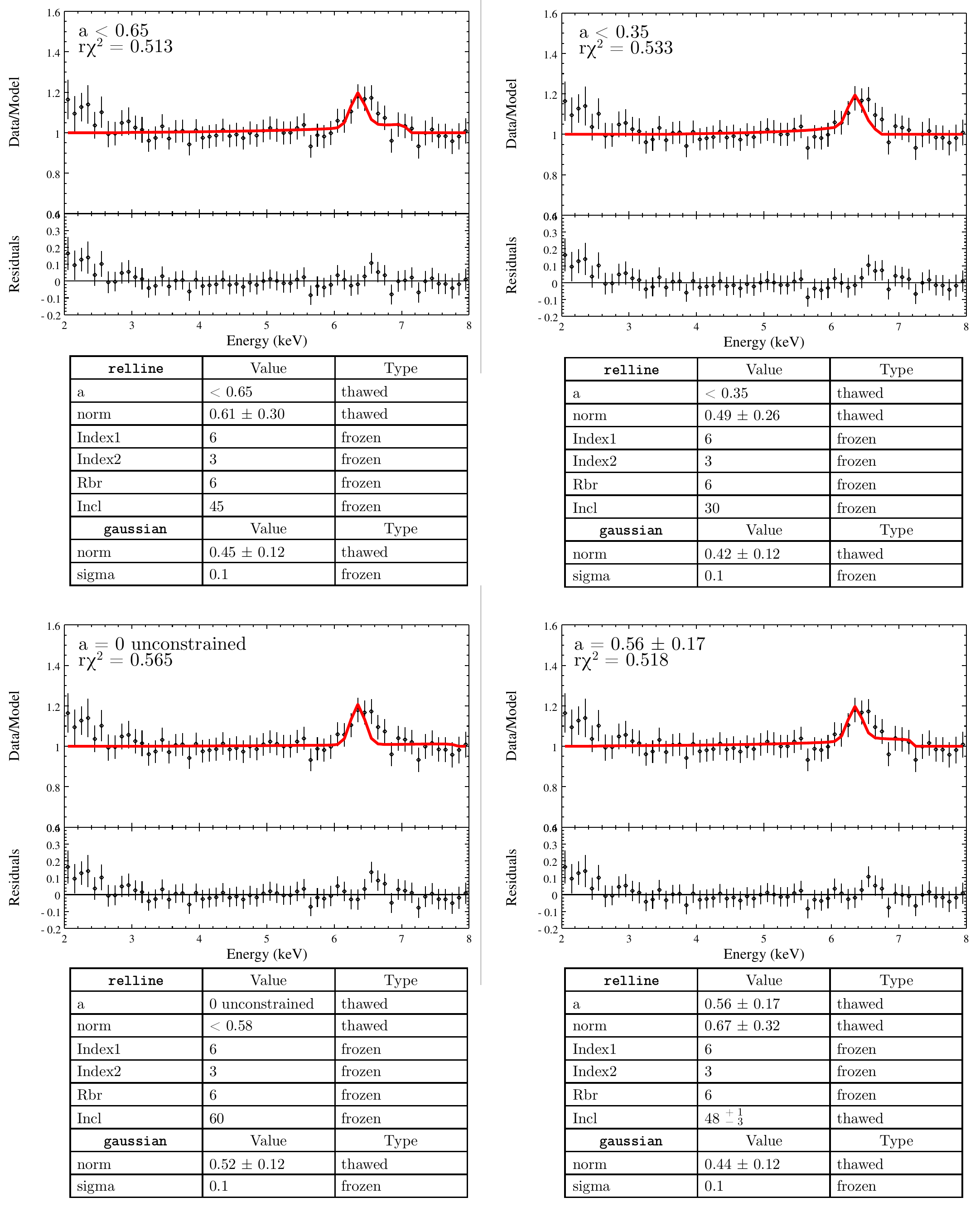}} \\
\caption{\texttt{relline}+\texttt{gaussian} model fits to the average data/model ratio for all CCLS sources. The \texttt{relline} and \texttt{gaussian} line energy are fixed at 6.4 keV, and the \texttt{gaussian} width is fixed to 0.1 keV (the size of our bins). All other fixed parameters are frozen at default values, unless otherwise indicated. The spin parameter (a) and the normalization of both the \texttt{relline} and \texttt{gaussian} are left free to vary while the inclination is fixed to: (top, left) 45 degrees; (top, right) 30 degrees; (bottom, left) 60 degrees. We also allow the inclination to vary (bottom, right). \label{fig:fit:rgincl}}
\end{center}
\end{figure*}

\begin{figure*}
\begin{center}
\resizebox{150mm}{!}{\includegraphics{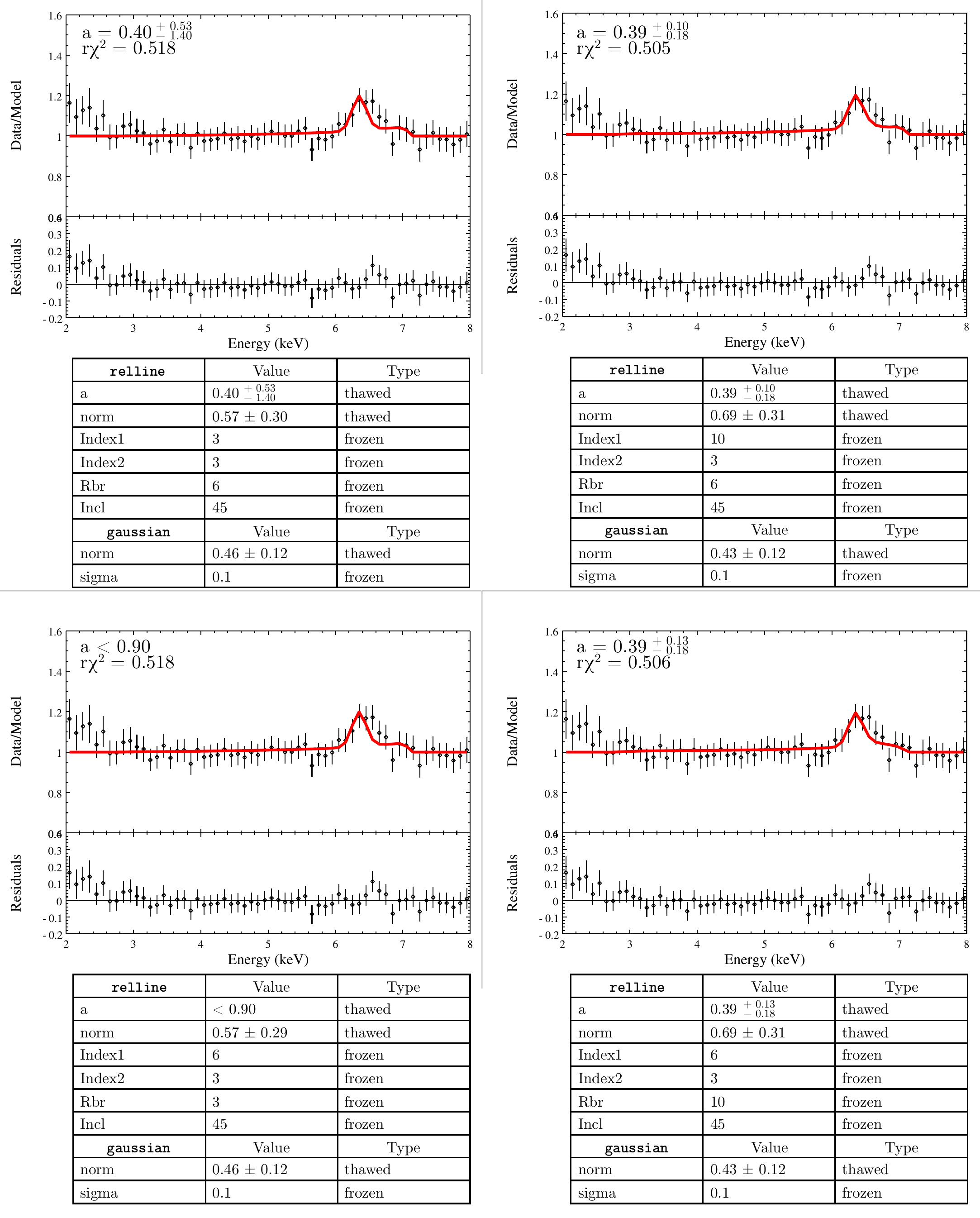}} \\
\caption{\texttt{relline}+\texttt{gaussian} model fits to the average data/model ratio for all CCLS sources. The \texttt{relline} and \texttt{gaussian} line energy are fixed at 6.4 keV, and the \texttt{gaussian} width is fixed to 0.1 keV (the size of our bins). All other fixed parameters are frozen at default values, unless otherwise indicated. The \texttt{relline} spin parameter (a) and normalization are left free to vary. We explore the parameter space of the emissivity index for r < Rbr (Index1) and the radius of emissivity change (Rbr) both above and below their typical values of 6: (top, left) Index1 $=3$; (top, right) Index $1=10$; (bottom, left) Rbr $=3$; (bottom, right) Rbr $=10$. \label{fig:fit:rgparam}}
\end{center}
\end{figure*}

\begin{figure*}
\begin{center}
\resizebox{150mm}{!}{\includegraphics{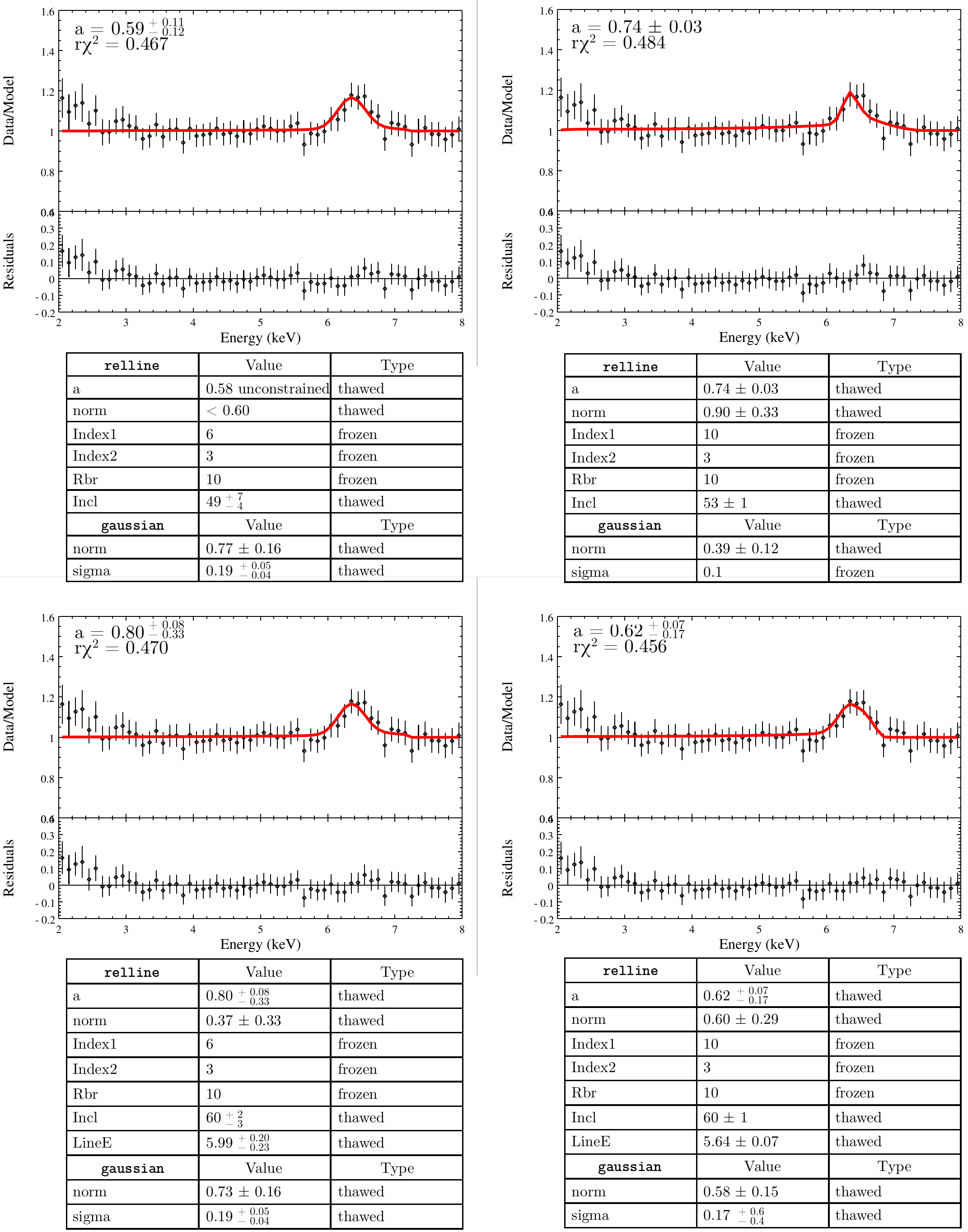}} \\
\caption{\texttt{relline}+\texttt{gaussian} model fits to the average data/model ratio for all CCLS sources. The \texttt{gaussian} line energy is fixed at 6.4 keV. All other fixed parameters are frozen at default values, unless otherwise indicated. The \texttt{relline} spin parameter (a), inclination, and both \texttt{relline} and \texttt{gaussian} normalizations are left free to vary. Building on previous fit results, we further explore the parameter space: (top, left) Index1 $=6$, Rbr $=10$, \texttt{gaussian} sigma left free to vary; (top, right) Index1 $=10$, Rbr $=10$; (bottom, left) Index1 $=6$, Rbr $=10$, \texttt{gaussian} sigma left free to vary, \texttt{relline} line energy left free to vary; (bottom, right) Index1 $=10$, Rbr $=10$, \texttt{gaussian} sigma left free to vary, \texttt{relline} line energy left free to vary. \label{fig:fit:rgbest}}
\end{center}
\end{figure*}

We have appended the data/model ratio fits for which we used \texttt{diskline} (Figure \ref{fig:fit:disk}), \texttt{relline} (Figure \ref{fig:fit:rlincl}, \ref{fig:fit:rlparam}), and the more complex \texttt{relline} + Gaussian models (Figure \ref{fig:fit:rgincl}, \ref{fig:fit:rgparam}, \ref{fig:fit:rgbest}) for the range of parameter space that was explored in our fits. 

\end{appendix}
\clearpage
\bibliography{bhs}


\end{document}